\documentclass{emulateapj}
\pdfoutput=1 
\usepackage{graphicx}

\slugcomment{Accepted for publication in The Astronomical Journal}

\shorttitle{IFU spectroscopy of HH 111}
\shortauthors{Cerqueira et al.}

\begin{document}

\title{Gemini-IFU spectroscopy of HH 111 \altaffilmark{1}}

\author{A. H. Cerqueira\altaffilmark{2}$^{,a}$, 
M. J. Vasconcelos\altaffilmark{2}$^{,a}$, A. C. Raga$^b$,
J. Feitosa\altaffilmark{3}$^{,a}$ and H. Plana$^a$}

\affil{$^a$LATO-DCET, Universidade Estadual de Santa Cruz\\ 
Rod. Jorge Amado, km 16, Ilh\'eus, BA, Brazil - CEP 45662-900} 

\email{hoth@uesc.br}

\affil{$^b$Instituto de Ciencias Nucleares, Universidad Nacional
Aut\'onoma de M\'exico, A.P. 70-543, 04510 D.F., M\'exico}

\altaffiltext{1}{Based on observations obtained at the Gemini
Observatory, which is operated by the Association of Universities
for Research in Astronomy, Inc., under a cooperative agreement with
the NSF on behalf of the Gemini partnership: the National Science
Foundation (United States), the Science and Technology Facilities
Council (United Kingdom), the National Research Council (Canada),
CONICYT (Chile), the Australian Research Council (Australia),
Minist\'{e}rio da Ci\^{e}ncia, Tecnologia e Inova\c{c}\~{a}o (Brazil)
and Ministerio de Ciencia, Tecnolog\'{i}a e Innovaci\'{o}n Productiva
(Argentina).}

\altaffiltext{2}{On a sabbatical leave at IPAG/UJF, Grenoble, France.}

\altaffiltext{3}{INCT-A scholarship}

\begin{abstract}

We present new optical observations of the HH 111 Herbig-Haro jet
using the Gemini Multi Object Spectrograph in its Integral Field
Unit mode. Eight fields of $5\arcsec$ x $3.5\arcsec$ have been
positioned along and across the HH 111 jet, covering the spatial
region from knot E to L in HH 111 (namely, knots E, F, G, H, J, K
and L). We present images and velocity channel maps for the [O I]
6300+6360, H$\alpha$, [N II] 6548+6583 and [S II] 6716+6730 lines,
as well as for the [S II]6716/6730 line ratio. We find that the
HH~111 jet has an inner region with lower excitation and higher
radial velocity, surrounded by a broader region of higher excitation
and lower radial velocity. Also, we find higher electron densities
at lower radial velocities. These results imply that the HH~111 jet
has a fast, axial region with lower velocity shocks surrounded by
a lower velocity sheath with higher velocity shocks.

\end{abstract}


\maketitle

\section{Introduction}

HH~111 (discovered by Reipurth 1989) is one of the two most remarkable
and better studied Herbig-Haro (HH) jets, the other one being HH~34
(shown to be a jet by Reipurth et al. 1986). The past observational
studies of HH 111 include:

\begin{itemize}

\item ground based (Reipurth et al. 1992; Podio et al. 2006) and
HST (Raga et al. 2002a) high and low resolution long-slit spectra,

\item optical (ground based: Reipurth et al. 1992; HST: Hartigan
et al. 2001) and IR (Coppin et al. 1998) proper motions,

\item evidence of multiplicity of the outflow source (Gredel \&
Reipurth 1993; Reipurth et al. 1999; Reipurth et al. 2000;
Noriega-Crespo et al. 2011),

\item discovery of an associated giant jet (Reipurth et al. 1997a),

\item detection of an associated, very well collimated molecular
outflow (Cernicharo \& Reipurth 1996; Nagar et al. 1997; Hatchell
et al. 1999; Lefloch et al. 2007)

\end{itemize}

Numerical simulations of variable jets calculated specifically for
modelling the observational properties of HH~111 were presented by
Masciadri et al. (2002) and Raga et al. (2002b).

Optically, the HH~111 system has a one-sided jet that appears at
$\sim 15''$ from an obscured source (detected at radio wavelengths
by Reipurth et al. 1999), extending to a distance of $\sim 40''$ W
from the source. However, a much more symmetric jet/counterjet
structure extending down to the position of the source is observed
in ground based (Gredel \& Reipurth 1994; Davis et al. 1994), HST
(Reipurth et al. 1999) and Spitzer (Noriega-Crespo et al. 2011) IR
images. At angular scales $>2'$ from the source, both of the outflow
lobes are detected optically, with total extent of $\sim 2^\circ$
for the whole system (see Reipurth et al. 1997a).

The kinematics of the HH~34 jet have been studied spectroscopically
(for a limited set of optical emission lines) with full spatial
coverage by Beck et al. (2007) and Rodr\'\i guez-Gonz\'alez et al.
(2012). Our present paper describes similar observations (i.e.,
high resolution spectroscopy with full spatial coverage), but for
the HH~111 jet.  In our Integral Field Unit (IFU) Gemini South
spectra the [S II] 6730, 6716; H$\alpha$; [N II] 6583, 6548 and [O
I] 6360, 6300 are detected, and have a high enough signal-to-noise
so that velocity channel maps can be generated.

The paper is organized as follows. In section 2 we describe the
observations and the data reduction. The results are described in
section 3. Finally, the work is summarized in section 4.

\section{Observations and data reduction}

\subsection{The observations}

HH~111 was observed at GEMINI North observatory on Oct 2007 and Nov
2010, using the GMOS instrument in IFU mode (Allington-Smith et al. 2002)
under the program GN-2007B-Q-9. We used the IFU in the
single slit mode with the R831\_G5302 grating, giving a $R= 4396$
resolution at 7570\AA. In the single slit mode, the IFU field of
view is $5\arcsec$ x $3\arcsec5$, each lens covering $0\arcsec2$
on the sky.

\begin{table}\label{Table 1}
\begin{center}
\caption{Target coordinates and offsets}
\label{tab1}
\begin{tabular}{ccccc}
\tableline
\tableline
Target &  RA & DEC  & $\Delta p$ & $\Delta q$\tablenotemark{a}\\

 & (h:m:s) & ($^{\circ}$:$^{\prime}$:$^{{\prime}{\prime}}$) & ($^{{\prime}{\prime}}$) & ($^{{\prime}{\prime}}$) \\

\tableline

IRAS 05491+0247\tablenotemark{b}  &
05:51:46.25     & 02:48:29.5  & - & -    \\
field 1  & 05:51:43.7    & 02:48:34.1  & - & -  \\
field 2 & & & 0 & 4.4 \\
field 3 & & & 0 & 8.8 \\
field 4 & & & 2.9 & 5.6 \\
field 5 & & & -2.9 & 5.6 \\
field 6 & & & 0 & -4.4 \\
field 7 & & & 0.214 & -9.07 \\
field 8 & & & 0.214 & -13.43 \\
\tableline
\end{tabular}
\tablenotetext{1}{The offsets $\Delta p$ and $\Delta q$ are given
in arcsec, and refers to displacements perpendicular ($p$) and
parallel ($q$) to the HH 111 PA; here defined as $\sim$ 83$^{\circ}$.
Field 1 is the reference to the displacement shifts.}
\tablenotetext{2}{Coordinates of the HH 111 IRS (Reipurth et al. 2000).}
\end{center}
\end{table}

The observations has made under exceptional seeing conditions.
Using the R-band images, the seeing estimates range from $0\arcsec44$
to $0\arcsec51$ for the first epoch observations (2007) and from
$0\arcsec55$ to $0\arcsec60$ for the second one (2010).

% Figure 1: cartoon

\begin{figure}
\centerline{
\includegraphics[width=7.5cm]{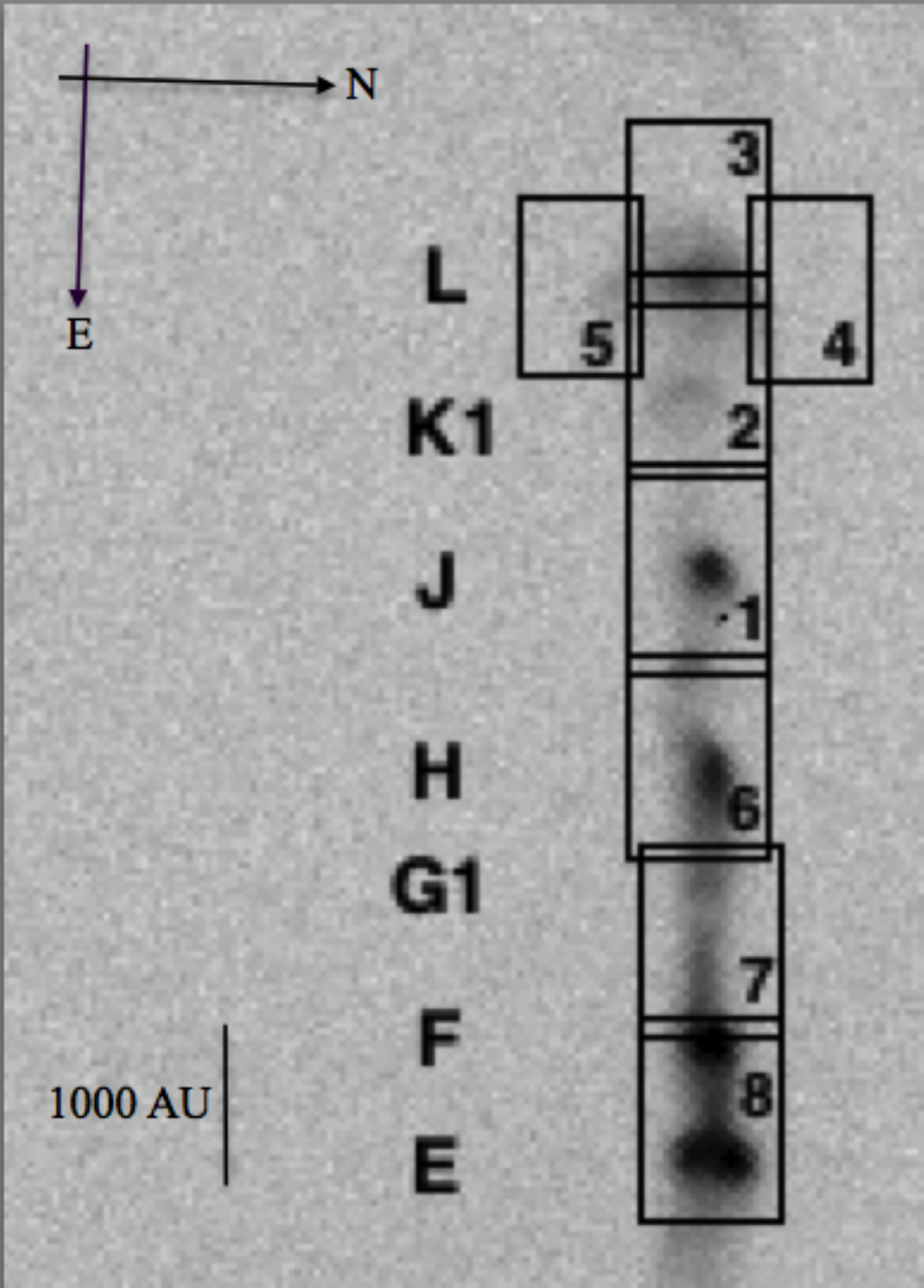}
}
\caption{Sketch of the eight fields observed with the GMOS-IFU
superimposed on the H$\alpha$ pre-image of HH 111 taken with the
Gemini North Telescope.  The observed fields are labeled by numbers
from 1 (centered on the brilliant knot J) to 8 (finishing the mosaic
in the double knot E).  The knots inside the HH 111 jet are labeled
by capital letters from E to L, following the nomenclature provided
by Reipurth (1989) and Raga et al. (2002a).  The offsets between each field
are indicated in Table \ref{tab1}. The N-E axes are indicated in the
Figure, as well as a distance scale.}
\label{fig1}
\end{figure}

\begin{figure}
\centerline{
\includegraphics[scale=0.32]{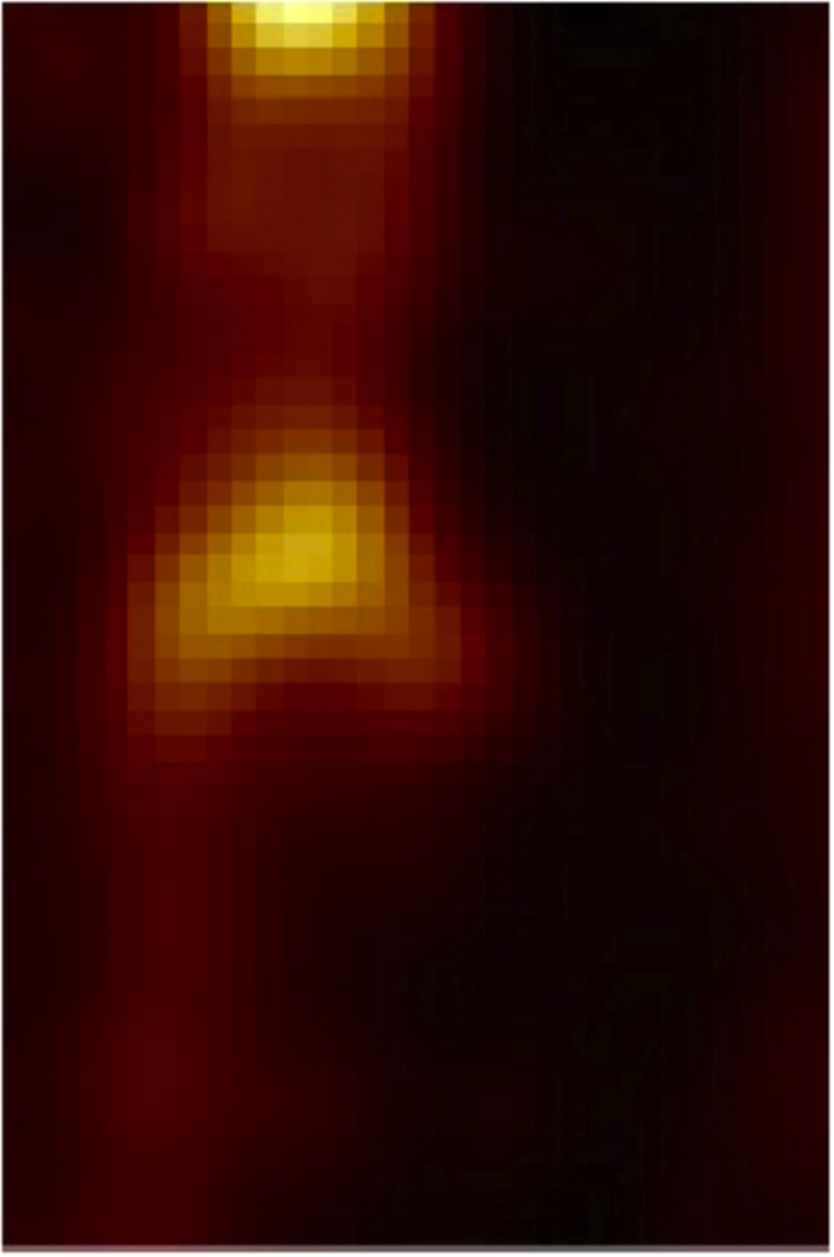}
\includegraphics[scale=0.32]{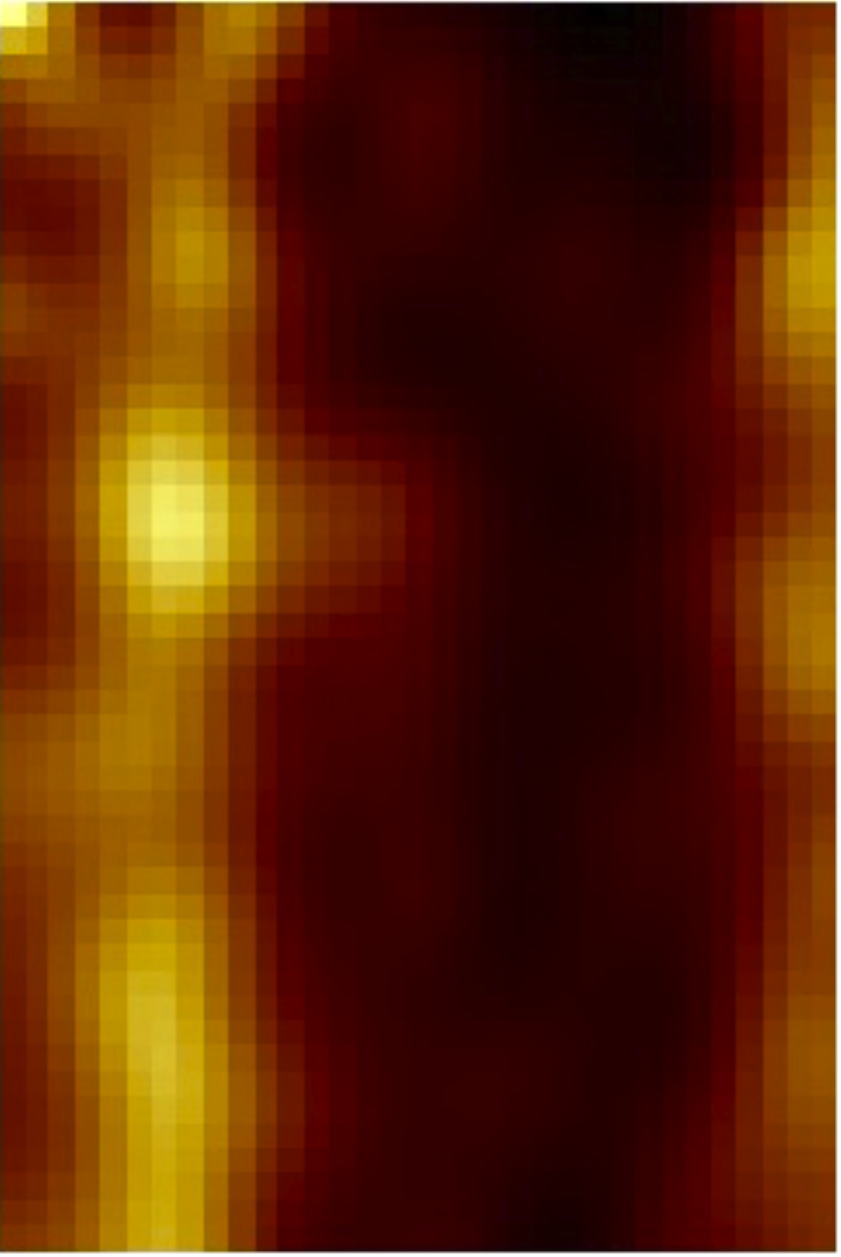}
\includegraphics[scale=0.32]{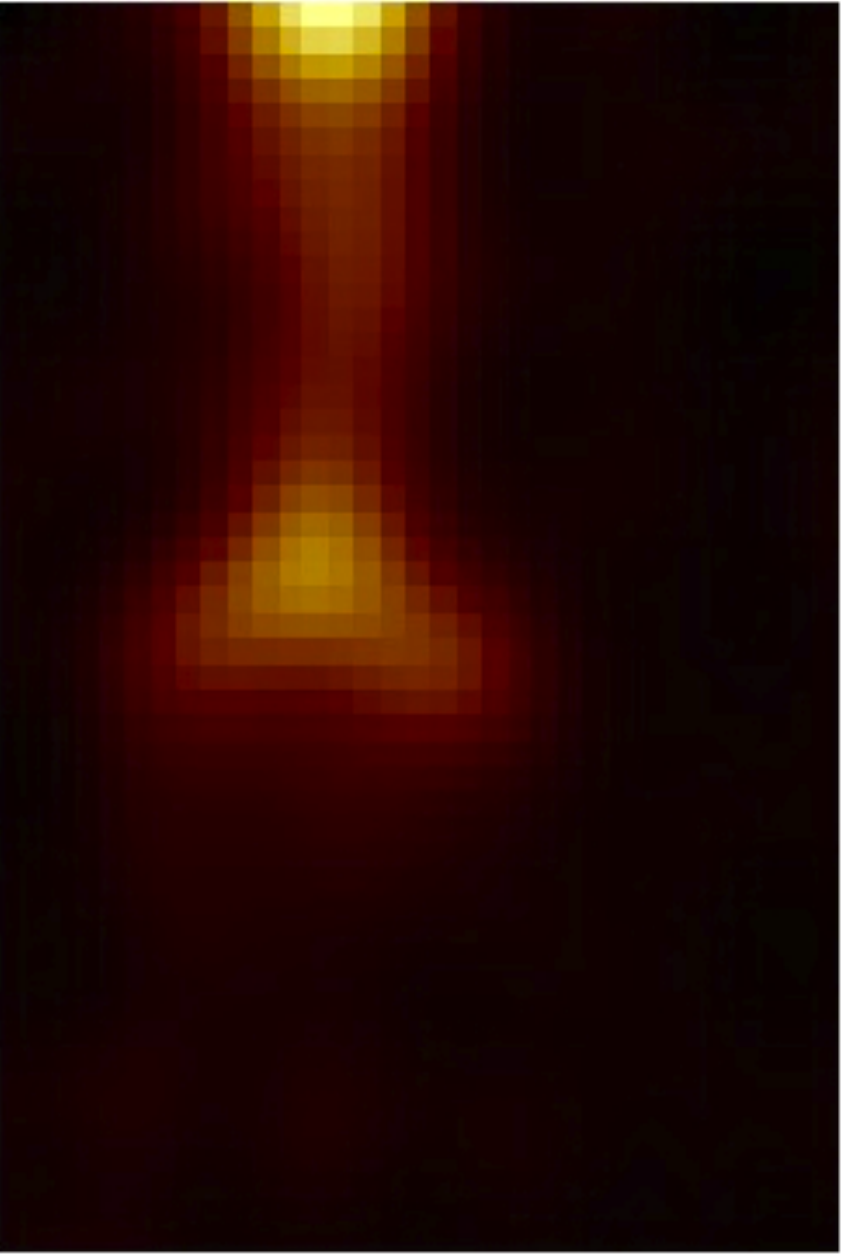}
}
\caption{{\it Left:} H$\alpha$ sub-cube of field 8 without any
filtering process ($3\arcsec5 \times 5\arcsec0$). {\it Middle:} The
sum of the first two eigenvectors, after PCA analysis of the data
cube in the spectral region defined by H$\alpha$, but excluding the
emission. {\it Right:} First image (left panel)
minus the low-frequency noise (middle panel). The results show that
the instrumental fingerprint has been removed from the original
data, without any introduction of artificial structures.}
\label{fig2}
\end{figure}

H$\alpha$ narrow-band filter pre-images (on and off) were also taken
(on August 22nd, 2007) to position the IFU field with higher accuracy.
In order to cover the target, we observed the 8 fields shown in
Figure \ref{fig1}. Each field was observed with a total exposure
of 1200s (three 400s exposures, which were then median averaged to
reduce the cosmic ray contamination). Fields 1 to 6 were observed
on October 19th 2007 and fields 7 and 8 on November 16th 2007.
Fields 1 to 6 were re-observed on November 2nd 2010, due to bad CCD
settings during the original observations.  Overlaps between the
different fields have been set to $0\arcsec4$ (corresponding to two
rows of IFU lenses).

Table \ref{tab1} gives the coordinates of the HH 111 IRS outflow
source (IRAS 05491+0247; see, Rodr\'{i}guez \& Reipurth 1994;
Rodr\'{i}guez et al. 1998; Reipurth et al. 2000) and the exact
offsets of the GMOS-IFU observed fields.  The offsets are in arcsec
and are taken with respect to the center of field 1 (center of knot
J, see Figure \ref{fig1}). The offsets are perpendicular ($\Delta
p$) and parallel ($\Delta q$) to the HH 111 outflow axis, which is
at a PA=$-83.94^{\circ}$ position angle (Raga et al. 2002a).

\subsection{Data reduction}

Data have been reduced using special reduction package provided
by Gemini Staff \footnote{http://www.gemini.edu/node/10795} using
the IRAF reduction package \footnote{Image Reduction and Analysis
Facility is a software developed by the National Optical Astronomy
Observatory - iraf.noao.edu}.  All raw images have been bias
corrected, trimmed and flat fielded. Flatfield images have also
been used to located the positions of the 1500 lenses on the frame.
Twilight images were used to estimate the grating response and after
the extraction of spectra for each lens, each one has been corrected
using this normalized response. Arcs, from the CuAr lamp, have also
been taken for the wavelength calibration. Using the bright OI night
sky line at 5577.338 \AA, we have estimated the wavelength calibration
accuracy to  0.1 \AA. The last step of the reduction was the sky
subtraction using the field located at 1\arcmin~ from the science
field. Finally the data cube has been created with the GFCUBE task
using a spatial resampling of 0.1\arcsec~ per pixel.  The total
spectral coverage goes from 4835.896 \AA~ to 6957.800 \AA. The
spectral sampling is 0.339 \AA~ per spectral pixel.
We then extracted sub-cubes near the H$\alpha$, [O I]$\lambda\lambda$
6300,6364, [N II]$\lambda\lambda$ 6548,6584 and [S II]$\lambda\lambda$
6716,6731 emission lines, for each the eight observed fields.

% Figure 3 [S II]6731 + 31

\begin{figure*}
\centerline{
\includegraphics[scale=0.5]{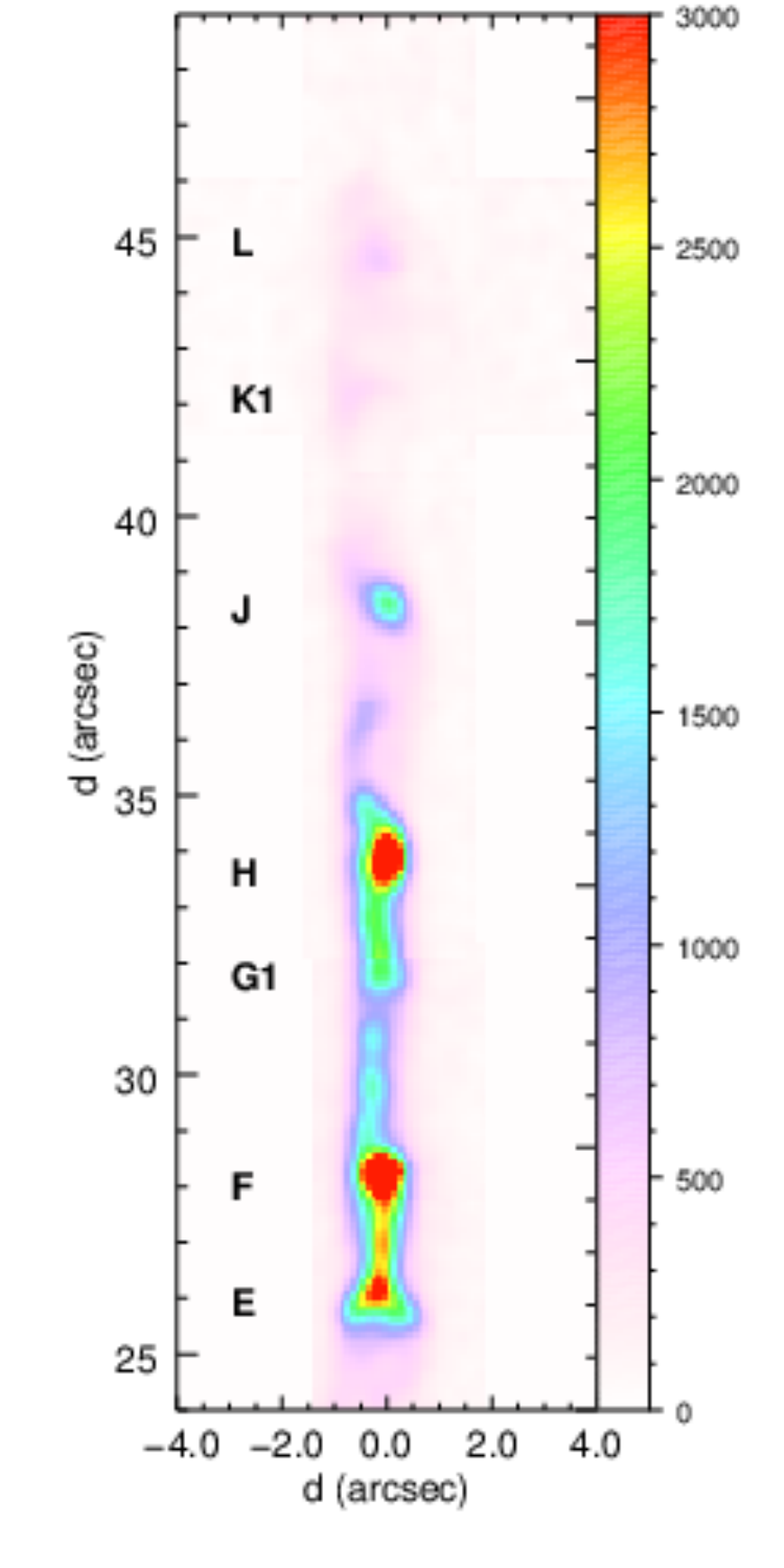}
\includegraphics[scale=0.5]{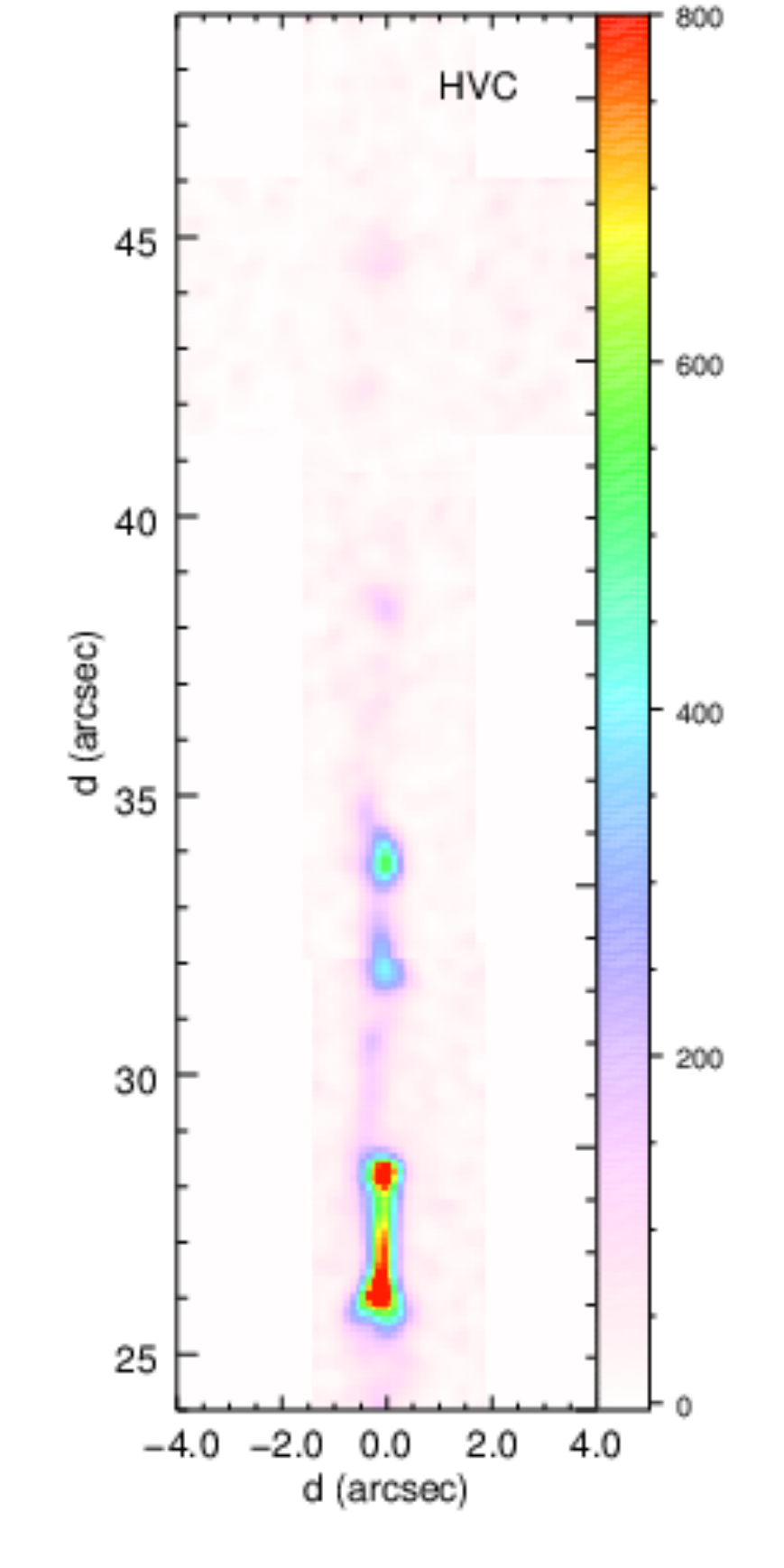}
\includegraphics[scale=0.5]{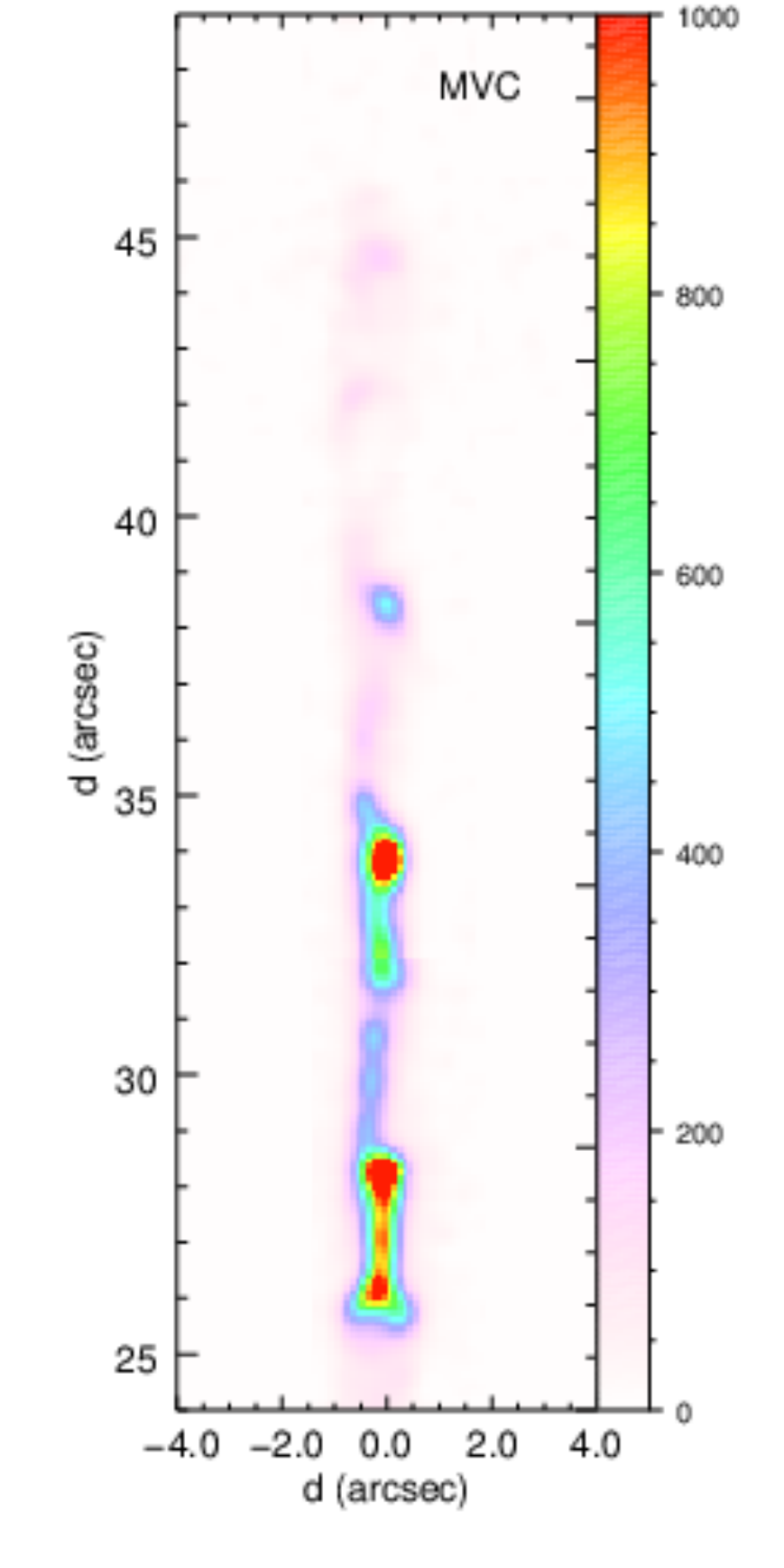}
\includegraphics[scale=0.5]{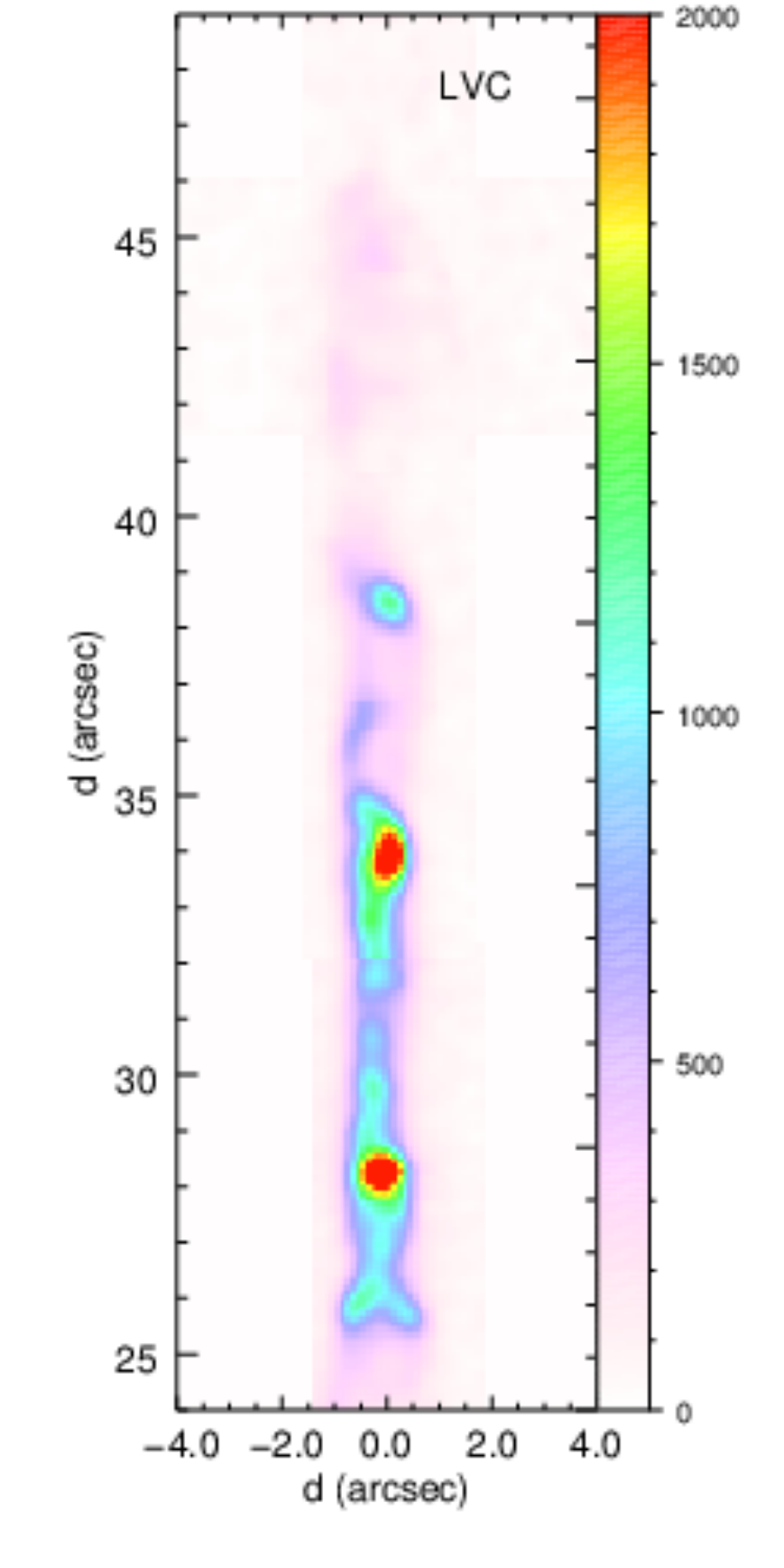}
}
\caption{{\it From left to right:} Mosaics of [S
II]$\lambda$6716+$\lambda$6731 for the integrated line profile
(left), and for three different radial velocities intervals: high
velocity (HVC, medium velocity (MVC) and low velocity (LVC) components.
The intervals are defined by: $v_{rad} < -100$ km s$^{-1}$ for HVC,
$-100 < v_{rad} < -70$ (in km s$^{-1}$) for the MVC  and $v_{rad}
> -70$ km s$^{-1}$ for the LVC. The coordinate axes are in arcsecs
(the origin coincides with the position of the driving source) and
the colorbar are in arbitrary units.}
\label{fig3}
\end{figure*}

% Figure 4 Halpha

\begin{figure*}
\centerline{
\includegraphics[scale=0.5]{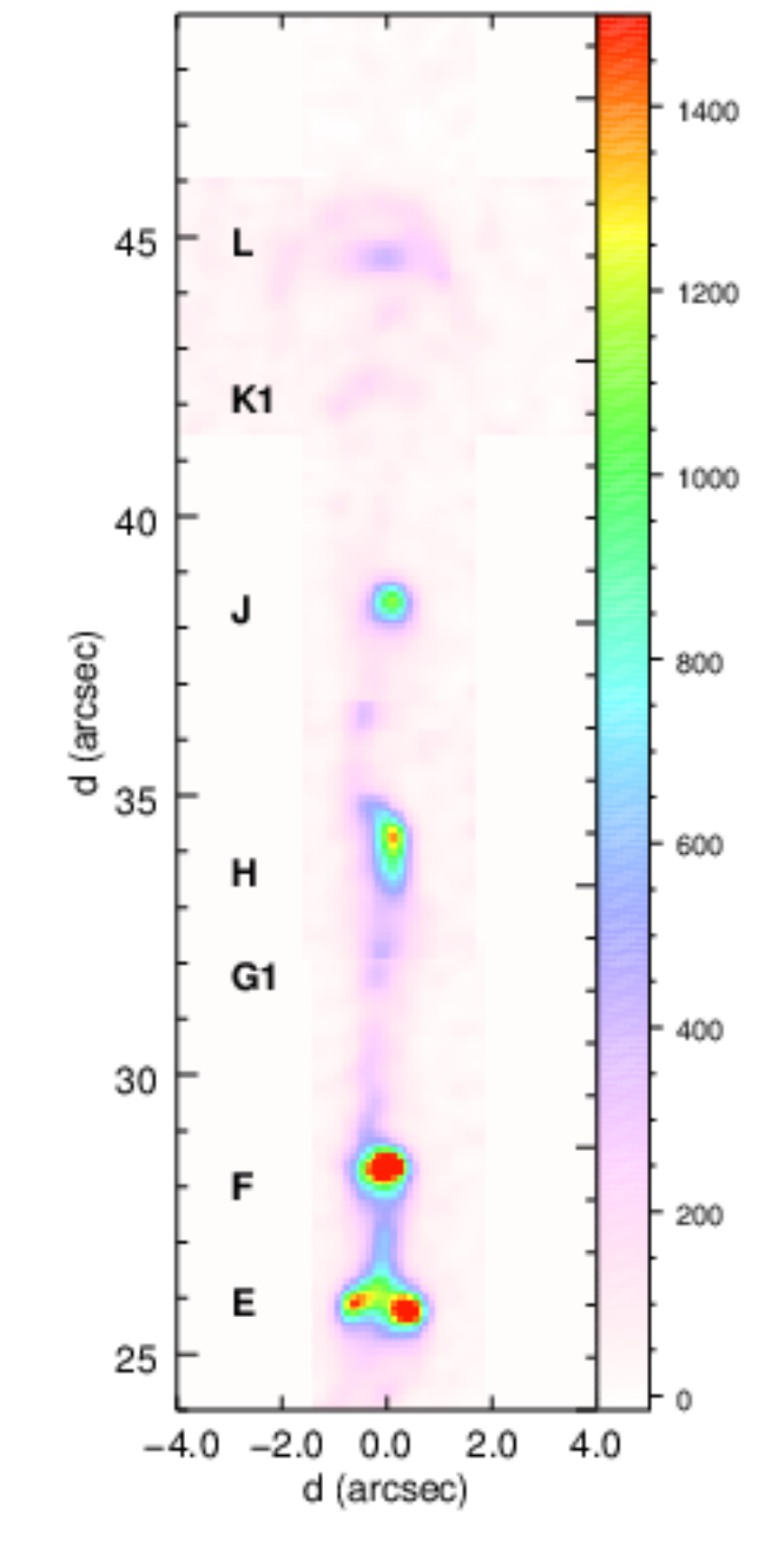}
\includegraphics[scale=0.5]{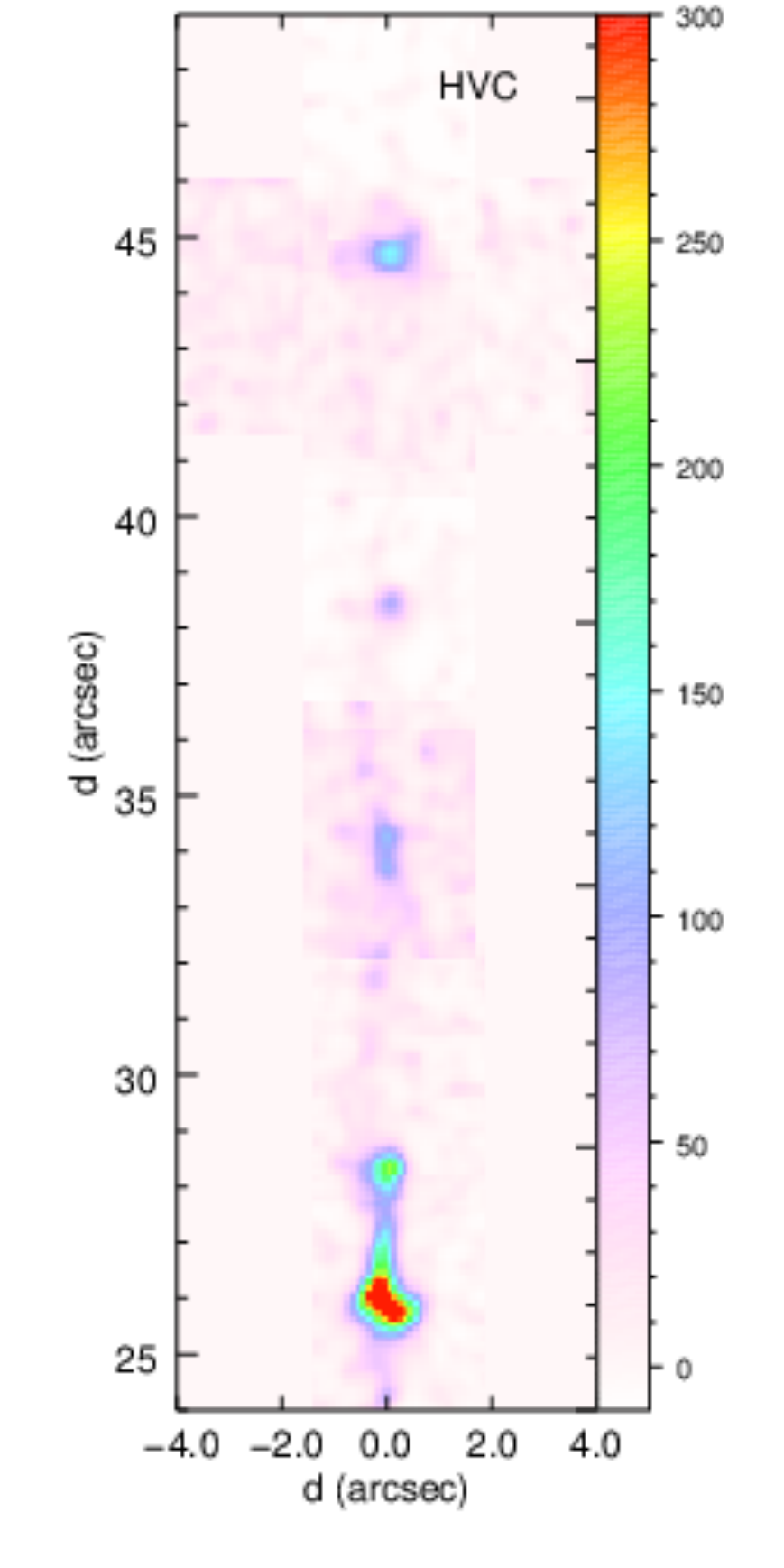}
\includegraphics[scale=0.5]{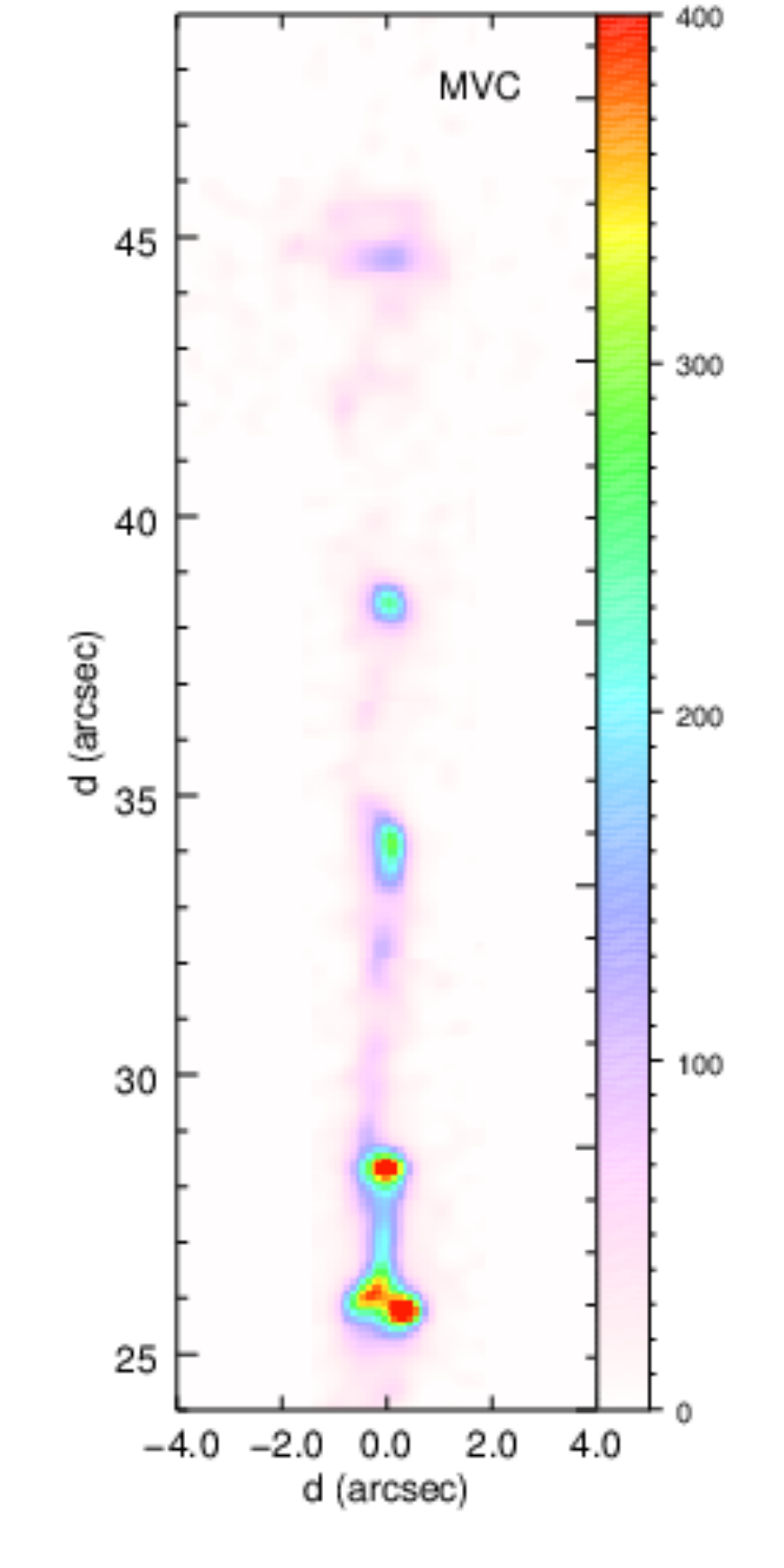}
\includegraphics[scale=0.5]{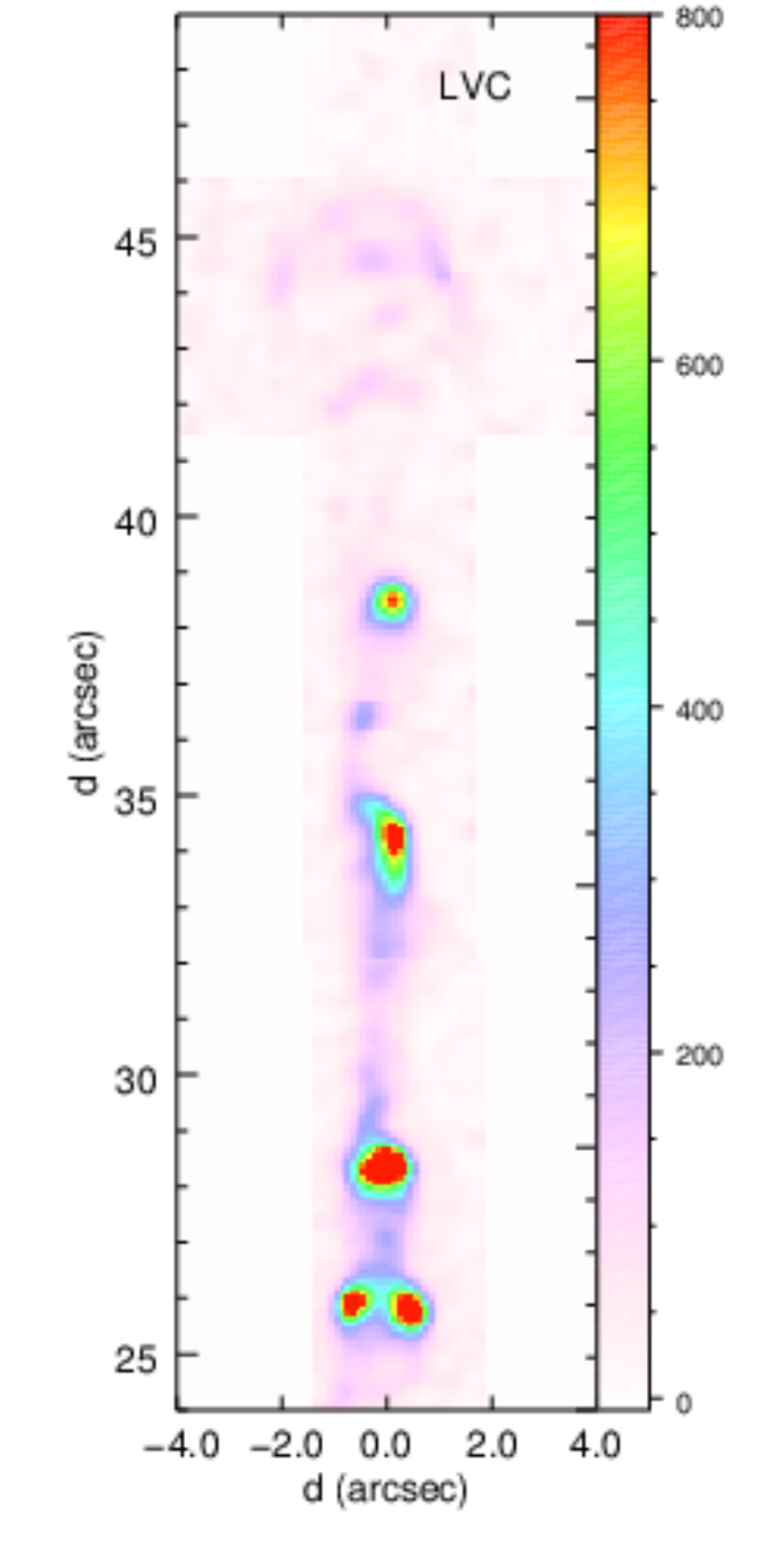}
}
\caption{The same as in Figure \ref{fig3} for H$\alpha$.
\label{fig4}}
\end{figure*}

\begin{figure*}
\centerline{
\includegraphics[scale=0.5]{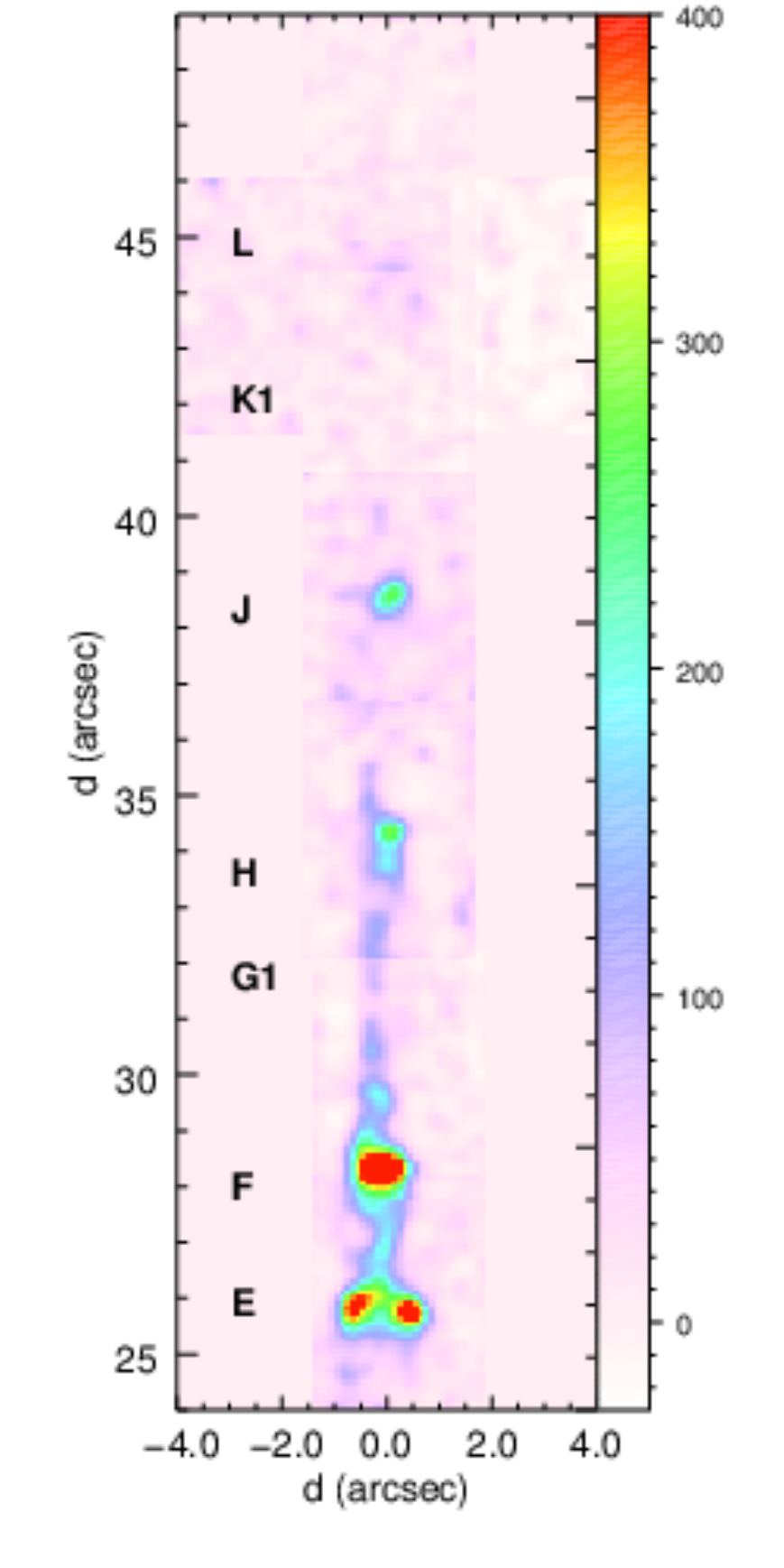}
\includegraphics[scale=0.5]{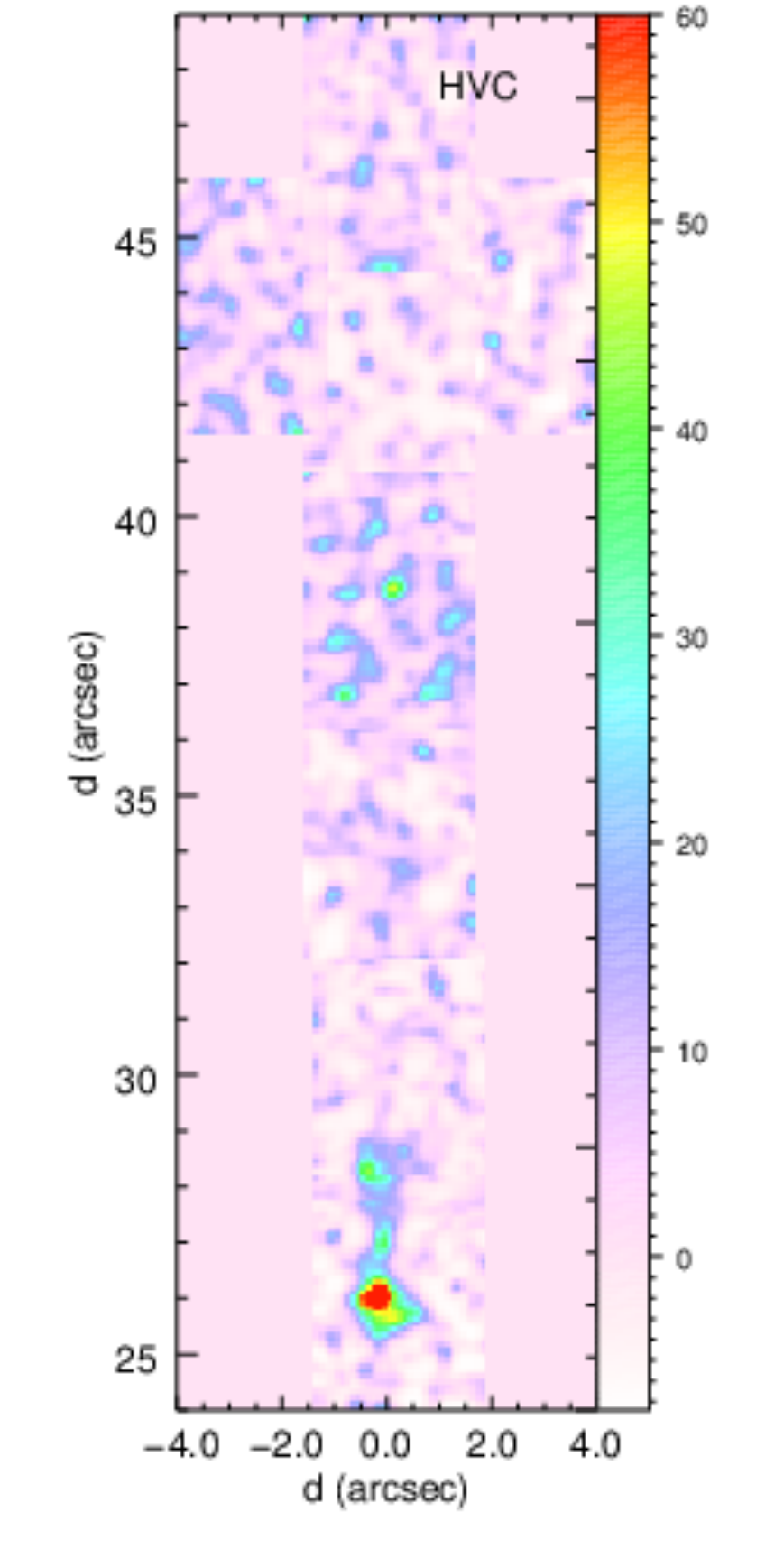}
\includegraphics[scale=0.5]{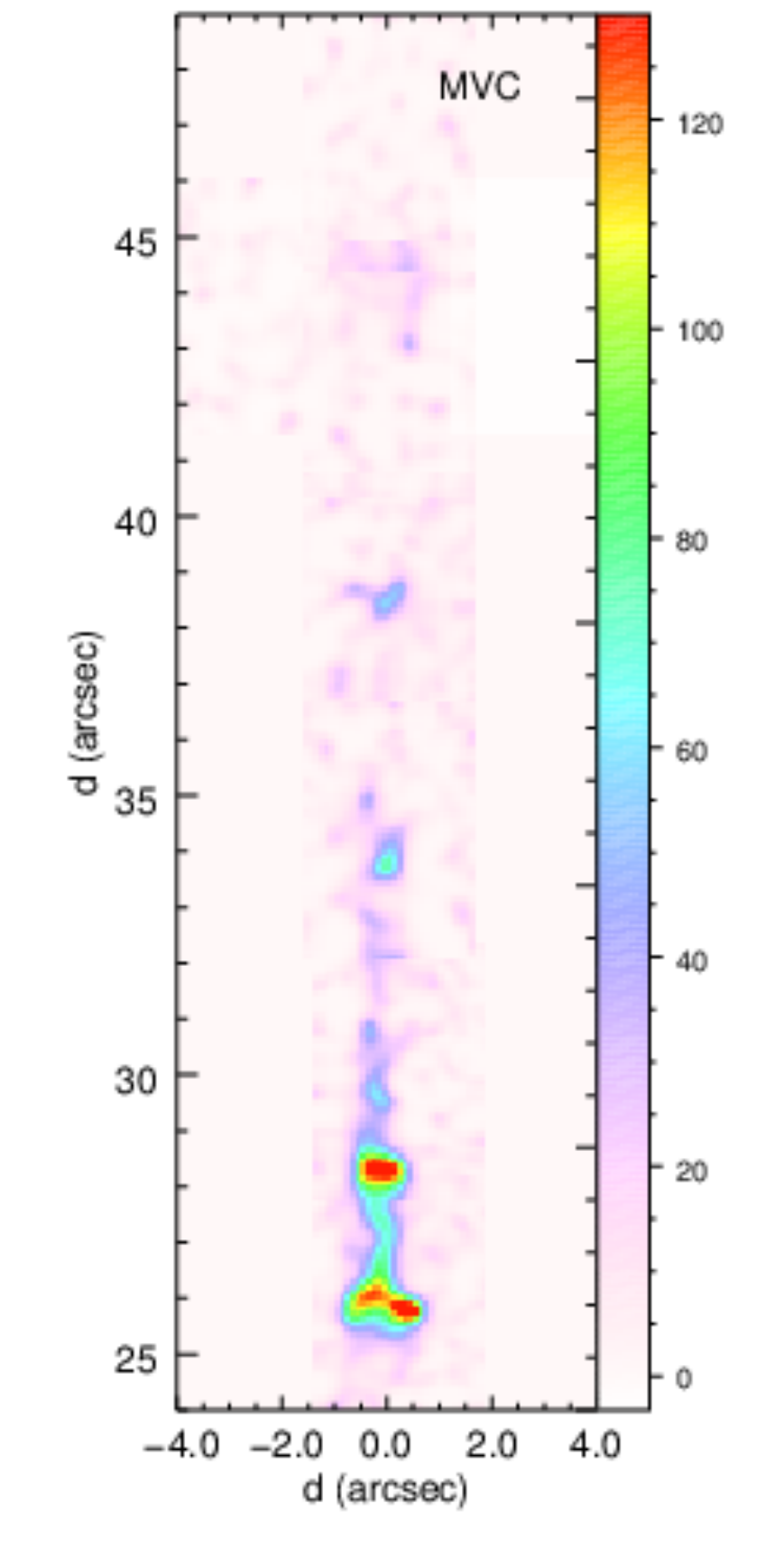}
\includegraphics[scale=0.5]{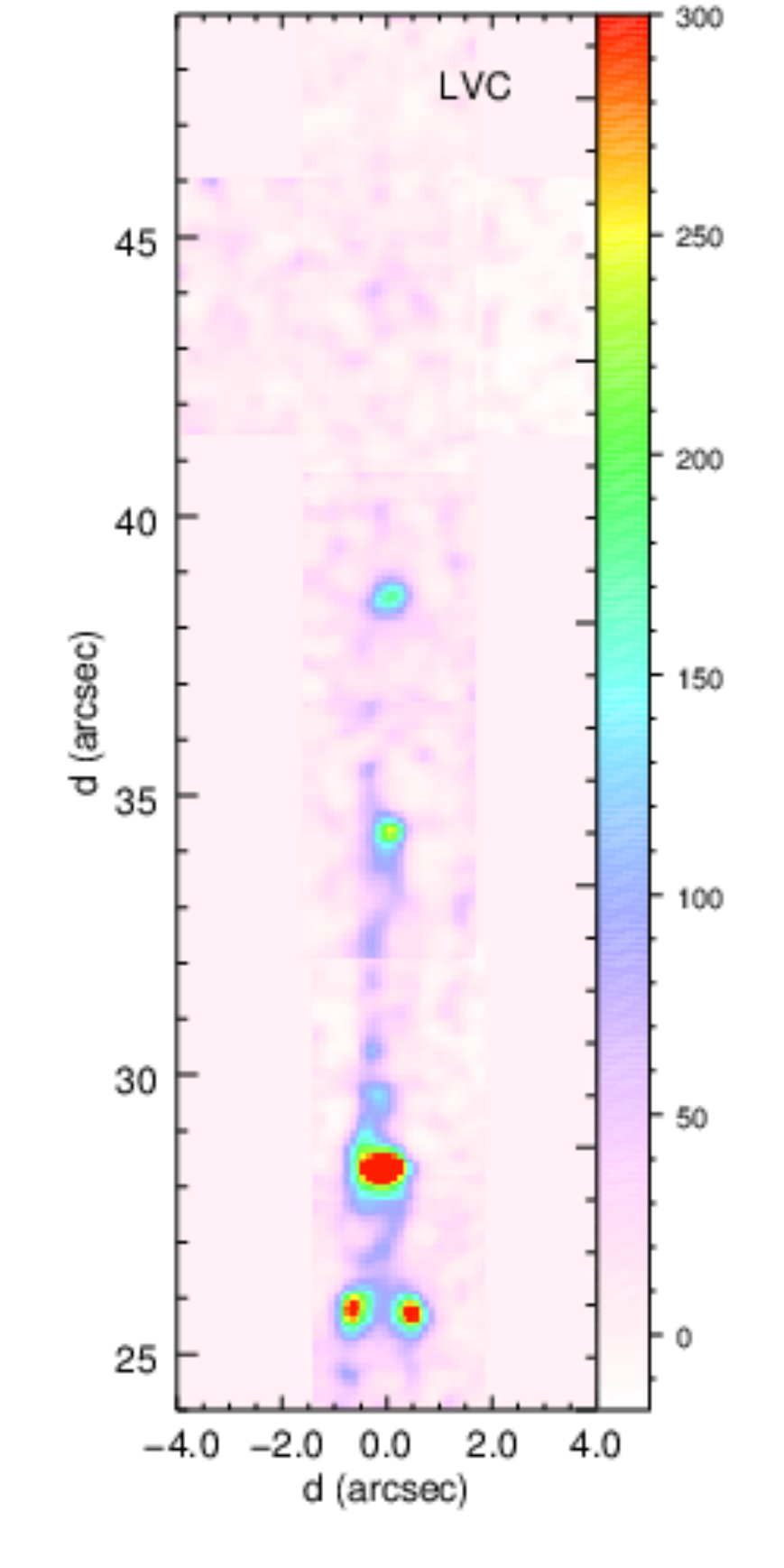}
}
\caption{The same as in Figure \ref{fig3} but for [N II]$\lambda$6548 +
$\lambda$6583.
\label{fig5}}
\end{figure*}

% Figure 6: [O I]

\begin{figure*}
\centerline{
\includegraphics[scale=0.5]{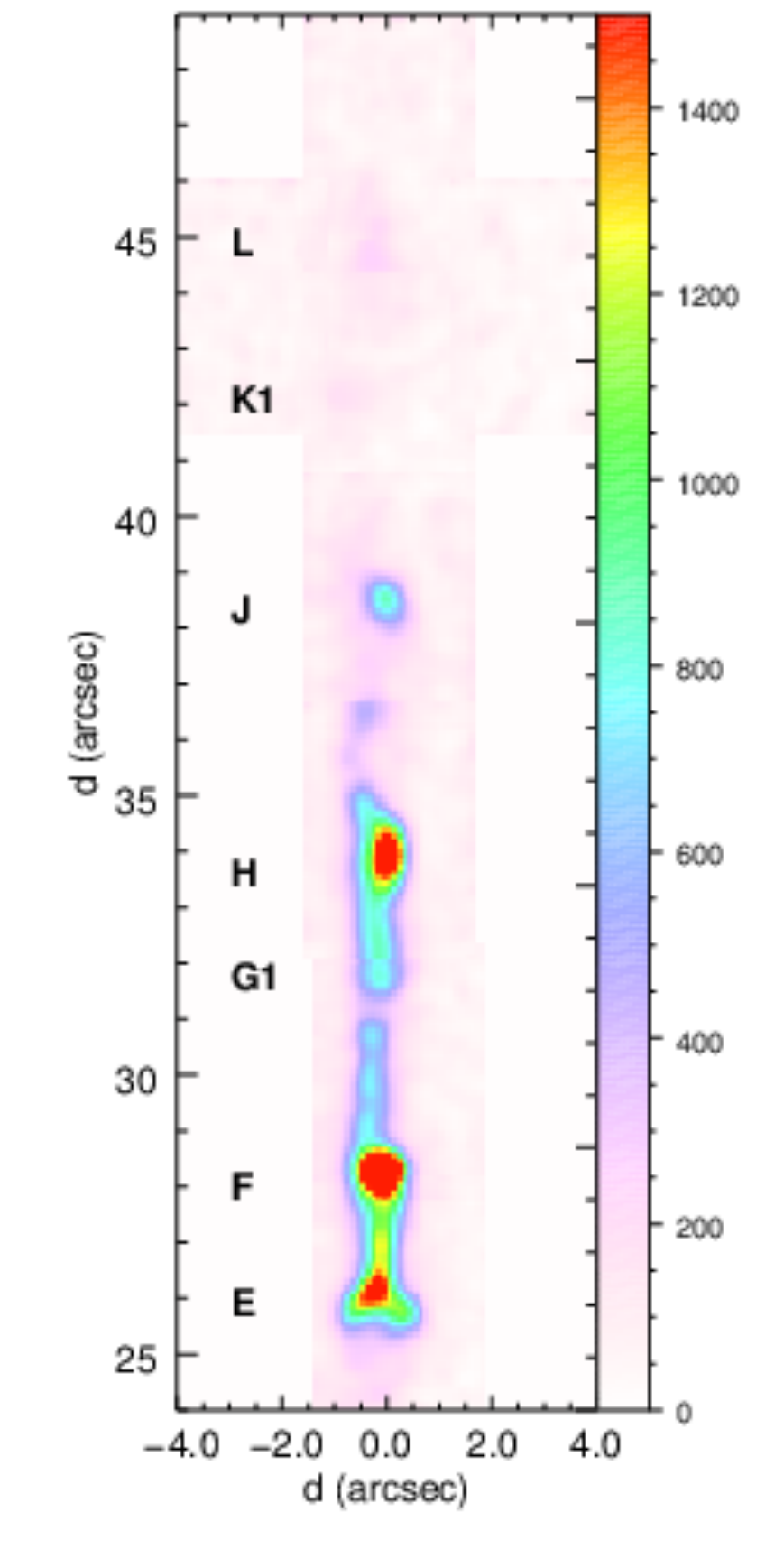}
\includegraphics[scale=0.5]{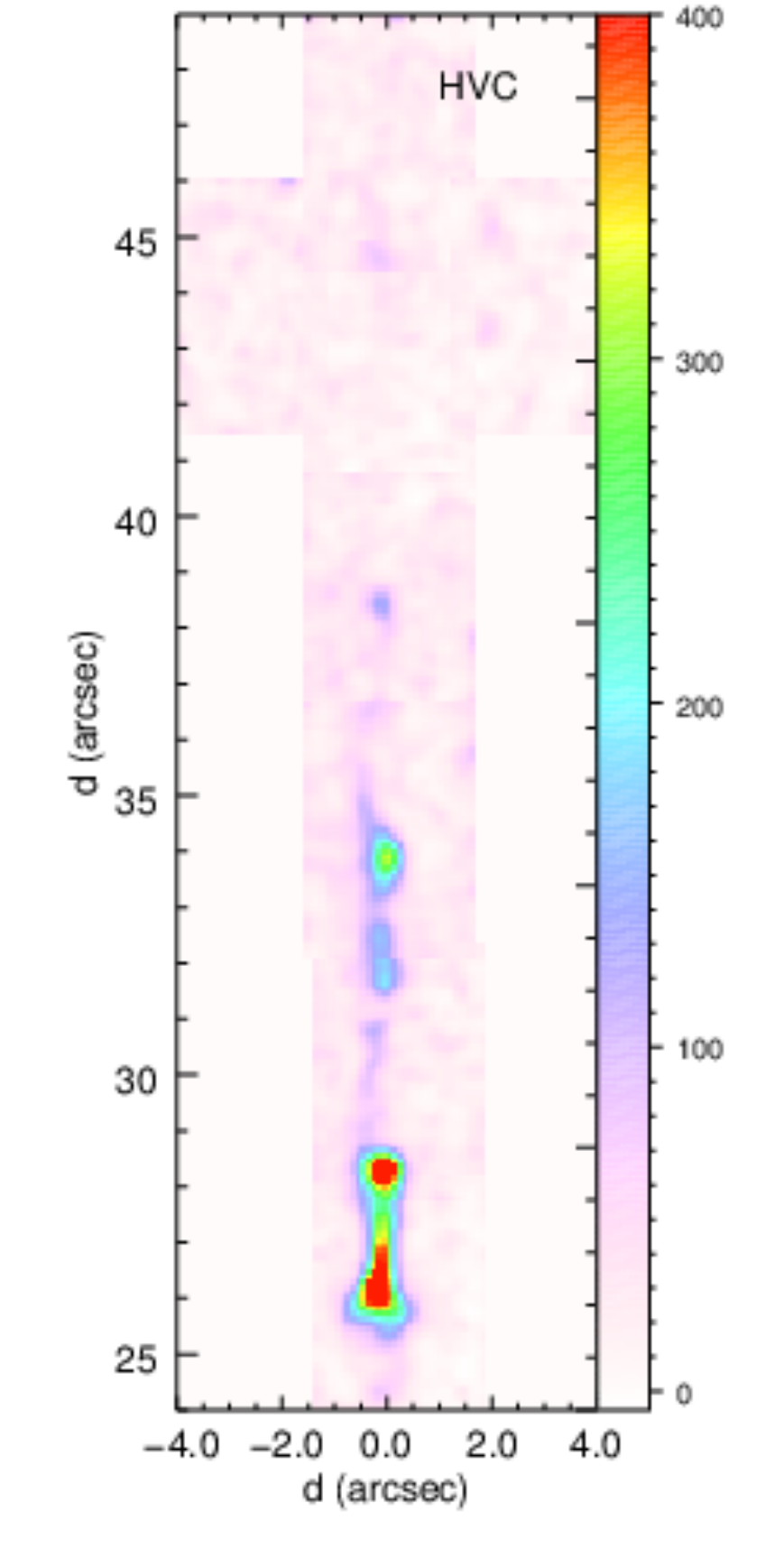}
\includegraphics[scale=0.5]{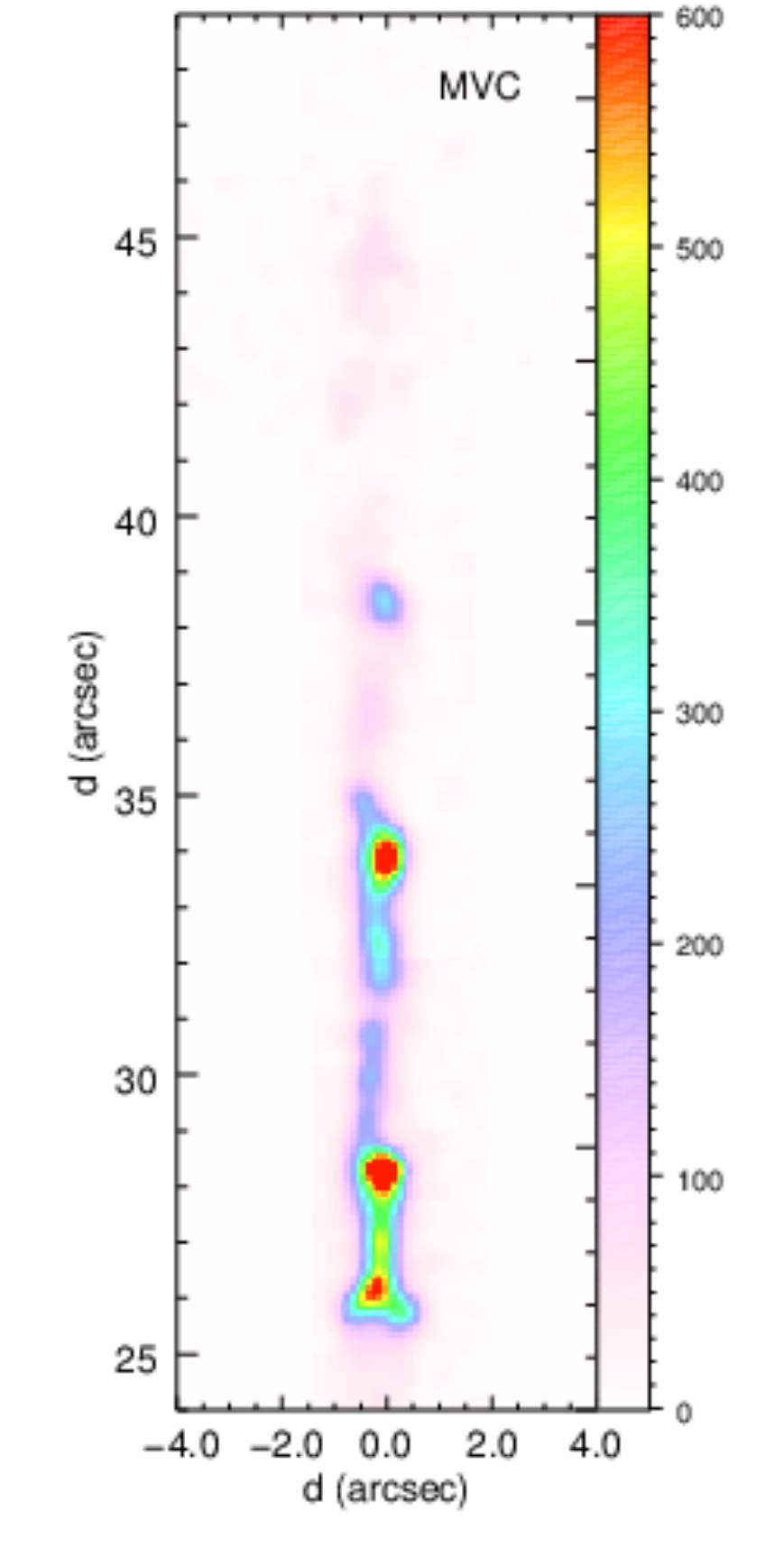}
\includegraphics[scale=0.5]{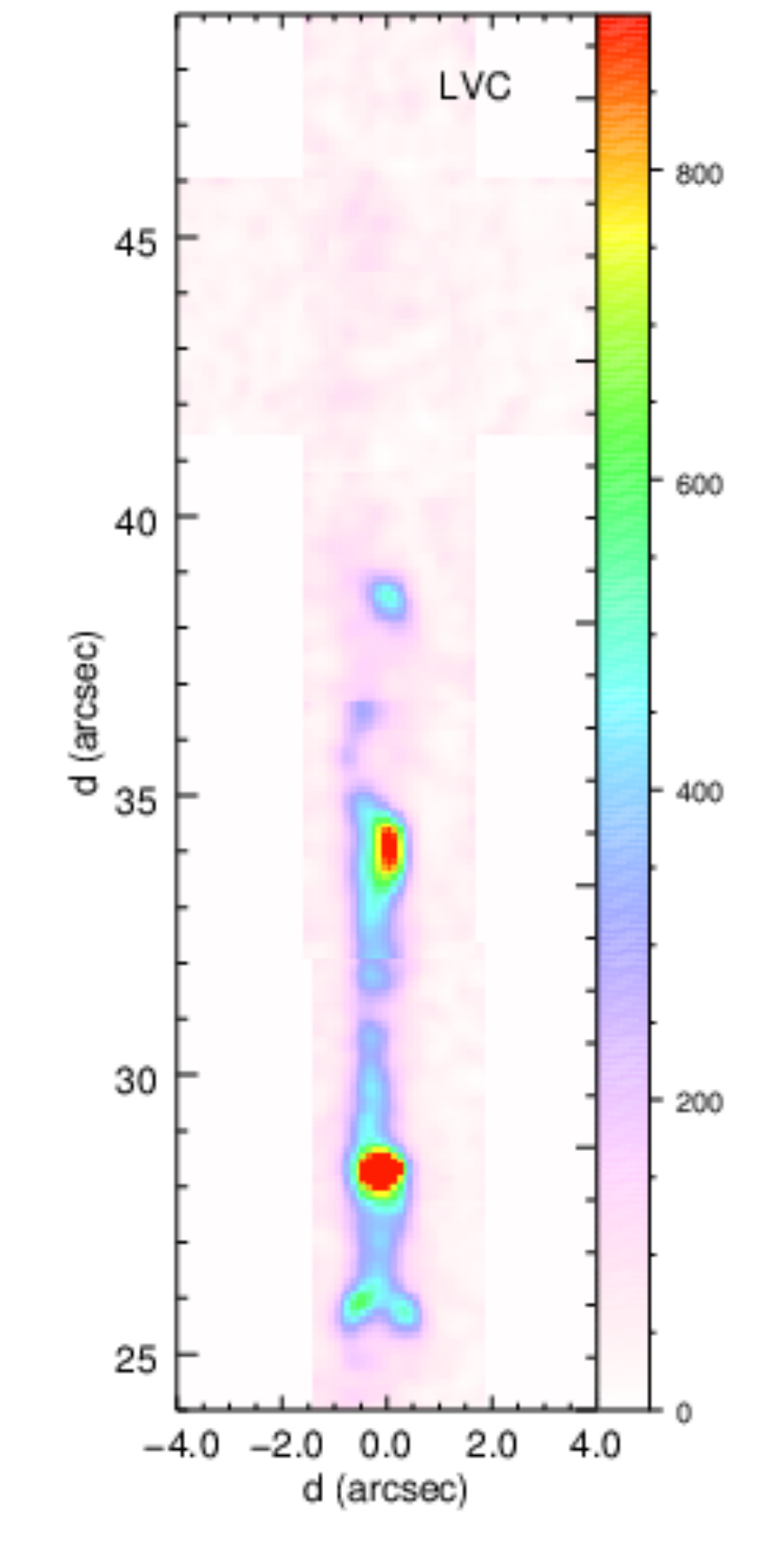}
}
\caption{The same as in Figure \ref{fig3} but for
[O I]$\lambda$6300 + $\lambda$6360.
\label{fig6}}
\end{figure*}

Further image treatment has been conducted using the pipeline
developed by Menezes et al. (2014; see also Ricci, Steiner \& Menezes
2011).  It consist in i) a spatial filtering process to remove
high-spatial frequency noise, ii) a Principal Component Analysis
(PCA) to remove instrumental spatial fingerprints and iii) a
Richardson-Lucy deconvolution process. In order to remove the high
spatial frequency noise, we first calculate discrete Fourier
transforms of the images resulting from each data cube, and apply
a low-pass band filter in the frequency space (Gonzalez \& Woods
2002). To this effect we follow the procedure described in Menezes,
Steiner \& Ricci (2014). In particular, we use a Butterworth $H(u,v)$
filter of order $n=6$ (see equation 7 in Menezes et al. 2014). An
inverse Fourier transform is then applied to the filtered data. We
should mention that similar noise filtering process has been already
applied in the context of long slit spectroscopy of HH jets (Raga
\& Mateo 1998).  The high-frequency cleaned data still show
low-frequency noise which we remove using the Principal Component
Analysis (PCA; see Ricci, Steiner \& Menezes 2011 and Steiner et
al. 2009 for a detailed discussion and method presentation). The
scope of this paper is not to present this technique.  However, it
is worth to mention briefly how it works.  The PCA technique allows
us to describe the data cube as a linear combination of an orthogonal
basis, which are the eigenvectors of a 2D covariance matrix. Such
a covariance matrix is obtained after scaling all the intensities
in the 3D data cube with the transformation:

\begin{equation}
\beta = \mu(i-1)+j
\end{equation}

% Figure 7: Ha LVC - HVC:

\begin{figure}
\centerline{
\includegraphics[scale=0.38]{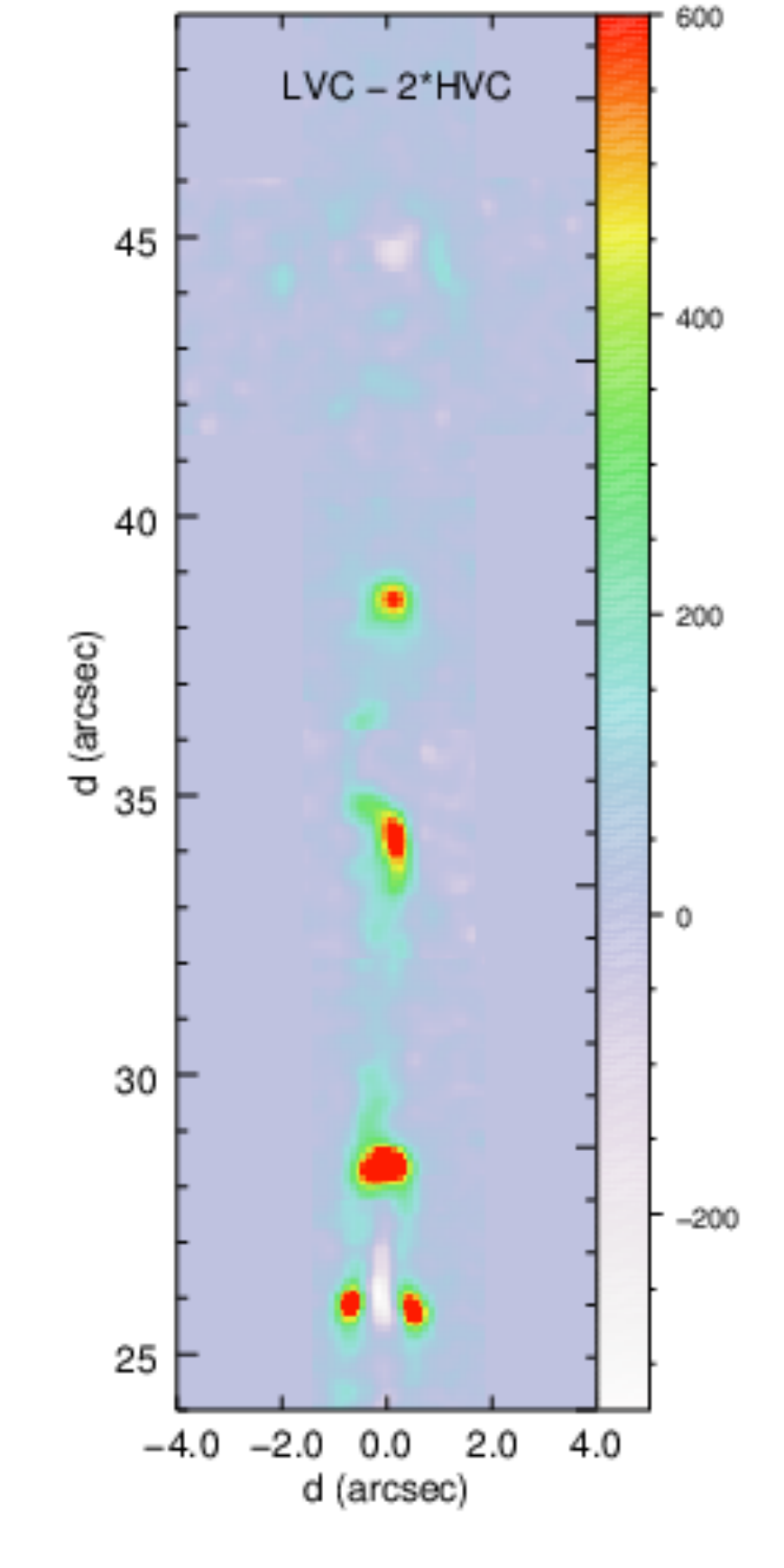}
\includegraphics[scale=0.38]{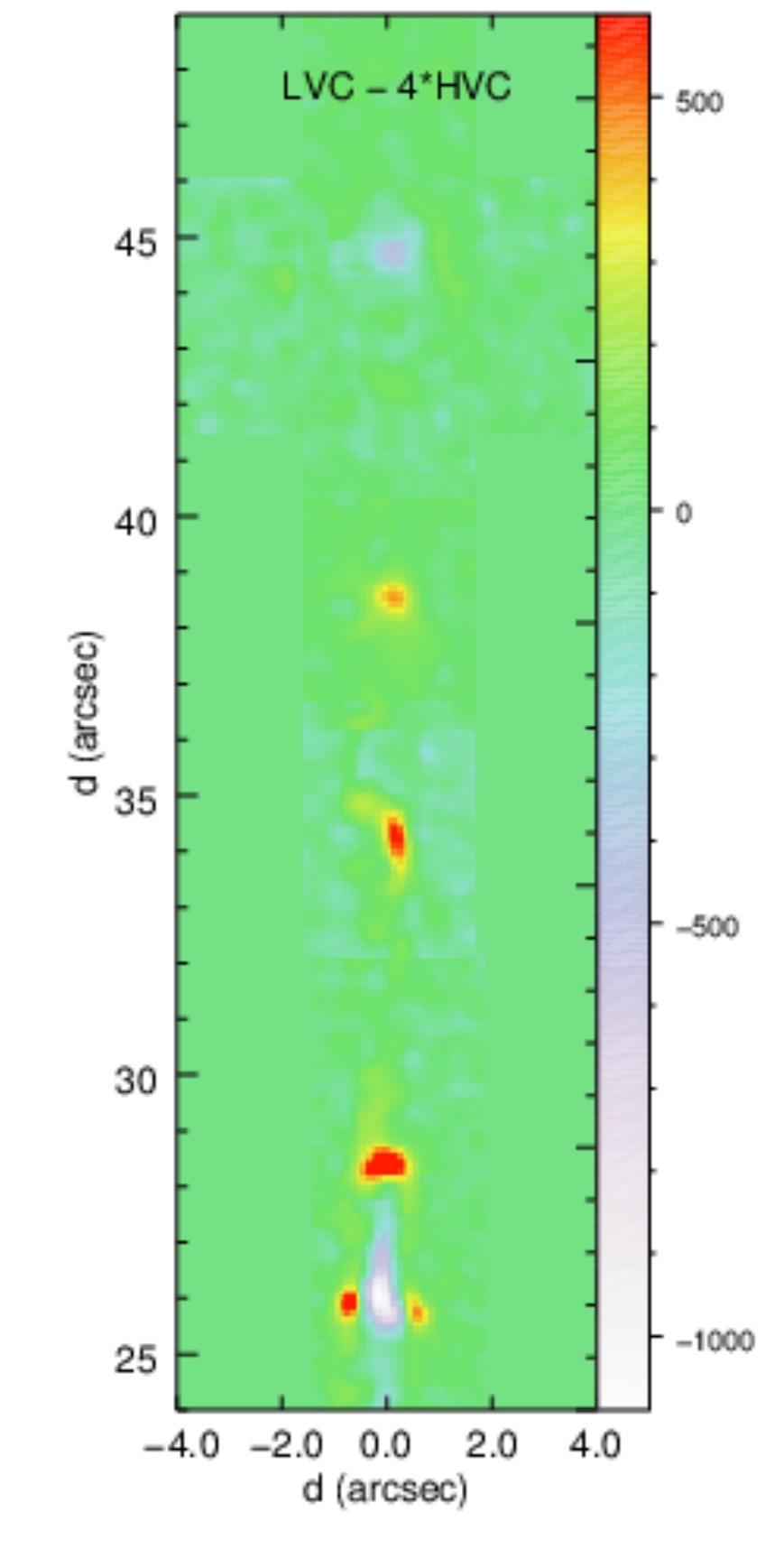}
}
\caption{Low-velocity and high-velocity components mosaic subtraction
in H$\alpha$ emission. {\it Left:} LVC - 2$\times$HVC. {\it Right:} LVC -
4$\times$HVC.
\label{fig7}}
\end{figure}

\noindent in a data cube with the spatial dimensions $n=\mu \times
\nu$: $1 \le i \le \mu$, $1 \le j \le \nu$. For each spaxel intensity,
$I_{ij\lambda}$, one defines a 2D intensity matrix {\bf I}$_{\beta
\lambda}$, which will be the subject of the PCA analysis. Its
covariance matrix is calculated, and the eigenvectors with their
associated variances are obtained (e.g., Steiner et al. 2009). The
higher its variance, the more representative of the data is a given
eigenvector.  As the orthogonality is ensured, we can compose
a tomogram, which is an image that represents the projection of the 
original data onto a selected eigenvector (or eigenvectors).  
In Figure \ref{fig2} we show a specific observed field
(field 8 in Figure \ref{fig1}), for a given emission line ($H\alpha$),
before any filtering process (a), after the filtering process but
extracting the emission line from data (b), and the (a)-(b) subtraction
in (c). We should mention that the tomogram in (b) was obtained by
adding the first two eigenvectors (without the emission line), which
we assume to correspond to the low-frequency noise, or the instrumental
fingerprint, as described in Ricci et al.(2011).

Finally, we have applied a Richardson-Lucy deconvolution procedure
with a Gaussian PSF, with a constant FWHM of $0\arcsec44$ for fields
7 and 8, and a constante FWHM of $0\arcsec55$ for the fields 1 to
6, as suggested by the seeing in our observations. Six interactions
has been used in order to deconvolve the original data with the PSF
(see equation 10 in Menezes et al. 2014). The whole procedure
(spatial re-sampling, low-noise filtering process, instrumental
fingerprints removal technique and Richardson-Lucy deconvolution)
is explained in detail in Menezes et al. (2014), who has developed
the technique\footnote{The routines are available electronically
at the following address:
\url{http://www.astro.iag.usp.br/$\sim$PCAtomography/}.}.

\section{Results}

\subsection{Images and velocity channel maps}

We present the spatially resolved line profiles as ``velocity channel
maps'' in three velocity intervals:

\begin{itemize}

\item HVC: a high (negative) velocity channel, with
radial velocities\footnote{The velocities has not been
corrected for the LSR velocity of $+ 23$ km s$^{-1}$
Reipurth 1989.} $v_{rad} < -100$ km s$^{-1}$,

\item MVC: a medium velocity channel, with $-100$ km s$^{-1}$ $<
v_{rad} < -70$ km s$^{-1}$,

\item LVC: a low velocity channel, with $v_{rad} > -70$ km s$^{-1}$.

\end{itemize}

We also present an image (for each spectral line) which consists
of an addition of these three velocity channel maps. The channel
maps and the images have been computed for mosaics composed of the
8 observed IFU fields (see Figure 1), and cover the emission of the
HH~111 jet at distances $\approx 24''\to 50''$ from the outflow
source.  The mosaics have been built taking into account the offsets
(given in Table \ref{tab1}).

In Figures \ref{fig3}, \ref{fig4}, \ref{fig5} and \ref{fig6} we
show the images and velocity channel maps for [S
II]$\lambda$6716+$\lambda$6731, H$\alpha$, [N
II]$\lambda$6548+$\lambda$6584 and [O I]$\lambda$6300+$\lambda$6360,
respectively. In the [S II] image (left frame of figure \ref{fig3})
we see knots E through L, which can be clearly recognized comparing
the image with the identifications of Reipurth et al. (1992) and
Reipurth et al. (1997b).

In the channel maps of all of the observed lines, it is clear that
there is a general trend of increasing jet width as a function of
decreasing velocity (i.e., in all of the observed emission lines
the jet becomes narrower as we go from the HVC to the MVC and to
the LVC, see Figures \ref{fig3}-\ref{fig6}). This result is in
qualitative agreement with the observations of Riera et al. (2001),
who found velocities decreasing away from the outflow axis in two
long-slit spectra cutting the HH~111 jet at the approximate positions
of knots D (not detected in the present observations) and F.

\subsection{Width vs. radial velocity dependence}

In order to illustrate the broadening of the jet at lower (i.e.,
less negative) radial velocities, in Figure 7 we show the LVC$-$HVC
channel map subtraction for the H$\alpha$ line.  This subtraction
map shows that at all positions along the observed region of the
HH~111, the high radial velocity emission is restricted to a narrow,
central region of the jet cross section.

\begin{figure*}
\centerline{
\includegraphics[scale=0.6]{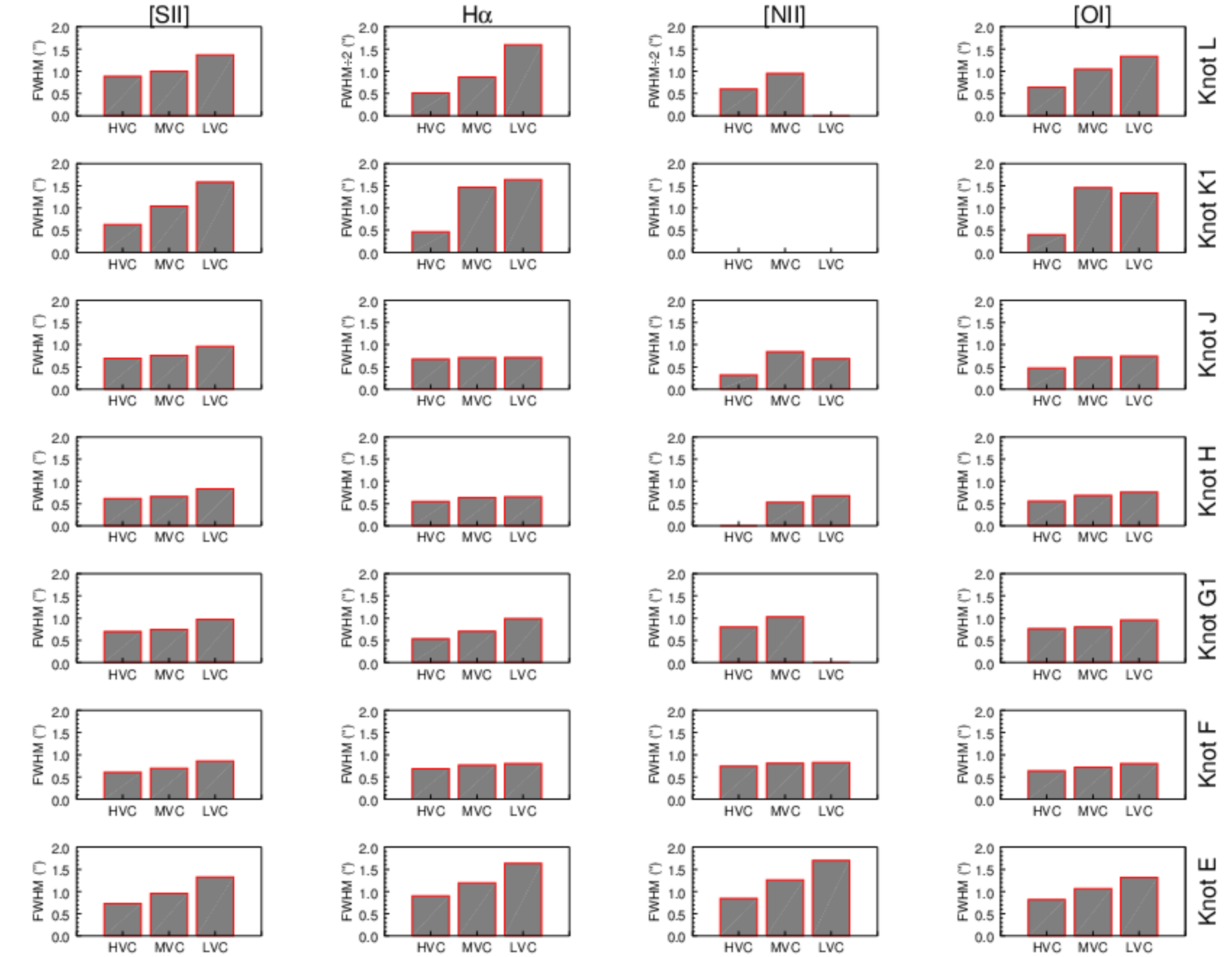}
}
\caption{FWHM (in arcsecs) of the emission on the HVC, MVC and LVC
channel maps for all of the emission lines (indicated on the top
of the figure), at the positions of knots E, F, G, H, J and L
(indicated on the right of the figure). The FWHM has been obtained
integrating the emission in bins of $1\arcsec0$ along the jet
(centered at the positions of the knots) and fitting a Gaussian to
each knot longitudinal profile. For [N II] line we found it difficult
to adjust a profile due to low S/N in some knots and for some
velocity components (see also Figure \ref{fig5}). For these cases,
the FWHM appears in the histogram as an zero.
The H$\alpha$ and [N II] histograms for
the knot L has the FWHM axis divided by a factor of 2.}
\label{fig8}
\end{figure*}

One can quantify the width vs. radial velocity dependence by measuring
the FWHM of the emission on the HVC, MVC and LVC channel maps. For
all of the emission lines, we measure the widths on the three channel
maps at the positions of knots E, F, G, H, J and L, integrating the
emission in bins of $1''$ along the jet bin (centered at the positions
of the knots).  The results of this procedure are shown in Figure
8, which shows the following features\footnote{The determination
of the FWHM for the knots G1, J, K1 and L, using the [N II] and [O
I] lines show high uncertainties due to their low S/N 
(see Figures \ref{fig5} and \ref{fig6}). In a few cases (knots
G1, H and K1; see Figure \ref{fig8}) we were not able
to determine the [N II] FWHM for some (if not all) velocity
components.}

\begin{itemize}

\item there is a systematic increase of width as a function of
decreasing radial velocity for all the knots, and in all emission
lines,

\item with the exception of the very broad knot E, there is a general
trend of increasing width vs. distance from the source,

\item in the lower excitation lines ([O I] and [S II]), the widths
are somewhat lower than in the higher excitation lines (H$\alpha$
and [N II]).

\end{itemize}

% Figure 8: widths of knots

This last effect has been noted by Reipurth et al. (1997), who note
that the H$\alpha$ emission is broader than the [S II] emission
(see their Figure 6, in which they present an H$\alpha$-[S II]
subtraction map). In our observations, we see that in general the [N II]
emission has widths comparable to the H$\alpha$ widths, and that
the [O I] emission has narrower widths, comparable to the ones of
the [S II] emission.

\subsection{The [S II] 6716/6731 line ratios}

\begin{figure*}
\centerline{
\includegraphics[scale=0.6]{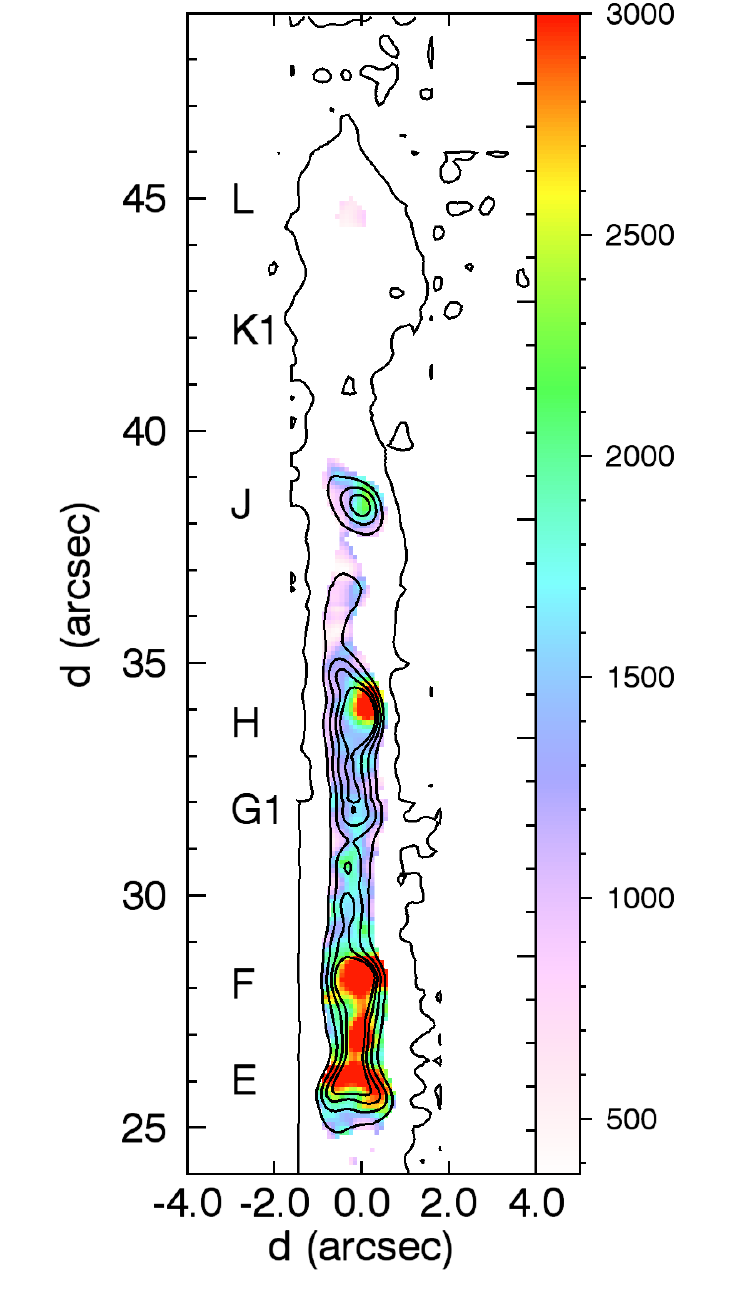}
\includegraphics[scale=0.6]{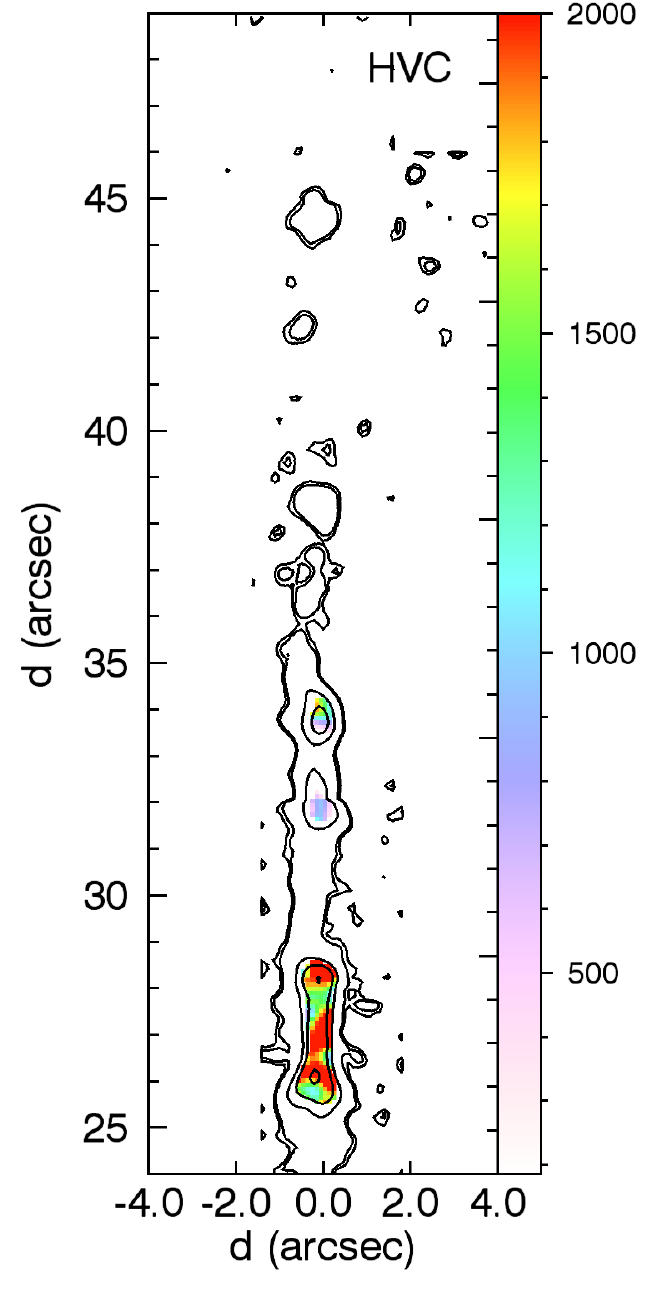}
\includegraphics[scale=0.6]{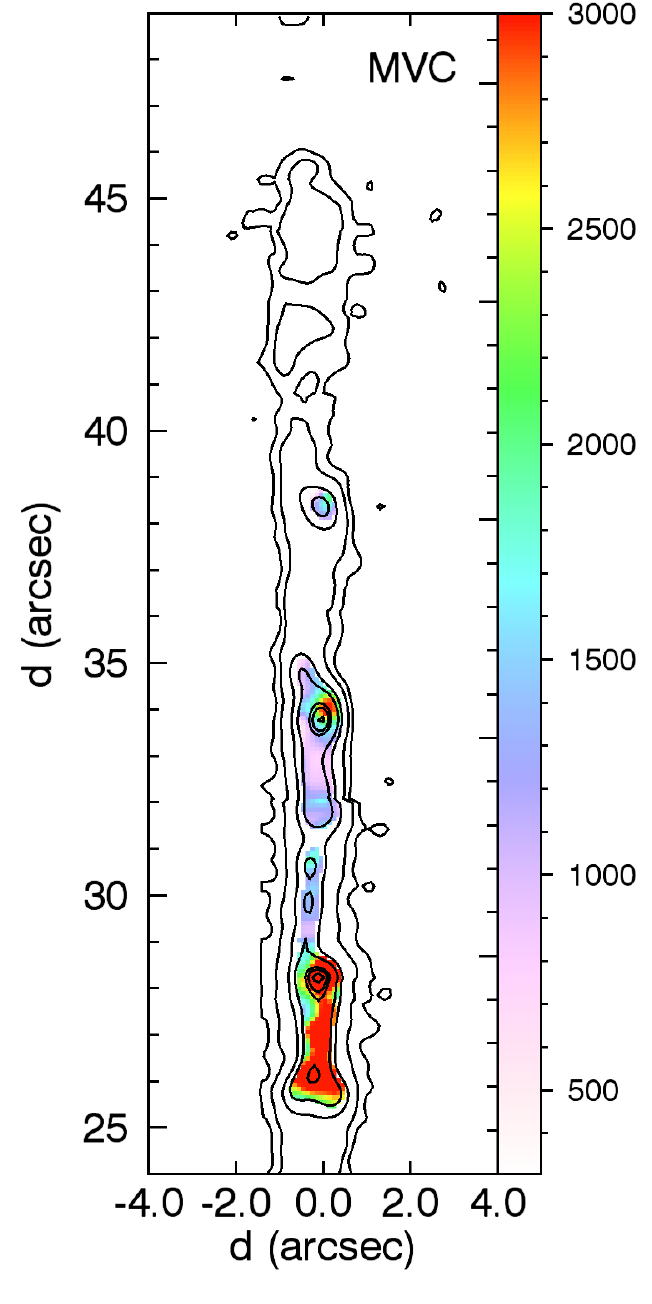}
\includegraphics[scale=0.6]{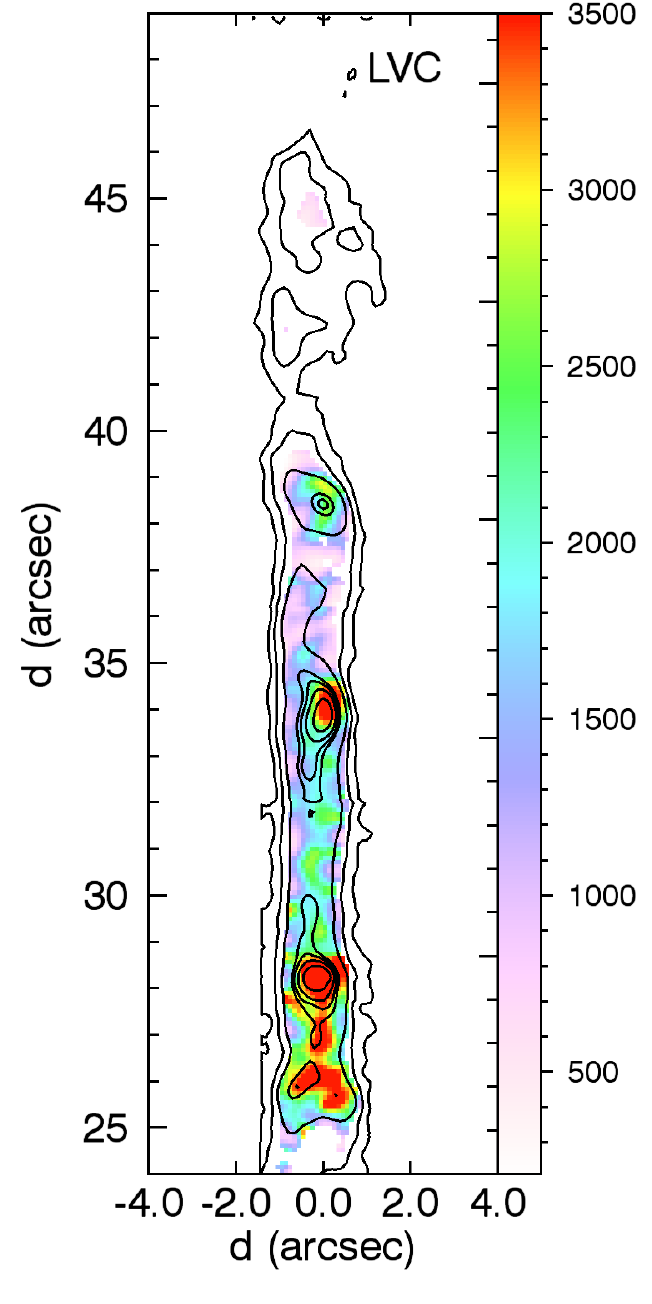}
}
\caption{{\it From left to right:} Mosaics of [S II]$\lambda$6716/[S
II]$\lambda$6731 line ratio for the integrated line profile, and
for three different radial velocities intervals: high velocity (HVC),
medium velocity (MVC) and low velocity (LVC) components. The intervals
are defined by: $v_{rad} < -100$ km s$^{-1}$ for HVC, $-100 < v_{rad}
< -70$ (in km s$^{-1}$) for the MVC  and $v_{rad} > -70$ km s$^{-1}$
for the LVC. The coordinate axes are in arcsecs
and the colorbar indicates the electron density in cm$^{-3}$.}
\label{fig9}
\end{figure*}

In Figure \ref{fig9} we present the spatial distribution of the [S
II]$\lambda$6716/[S II]$\lambda$6731 line ratio for the integrated
line profiles and for the HVC, MVC and LVC maps. This ratio gives
the electron density ($n_e$) of the medium
Osterbrock (1989). For calculating $n_e$ we assume a temperature of
10$^4$ K, consistent with the temperature of  $\sim 1.3\times10^4$
K found by Podio et al. (2006) for the HH~111 jet.

Superimposed on the spatial distribution of the electron density
are the (logarithmic spaced) iso-contours of the [S II]$\lambda$6716
+ $\lambda$6731 integrated intensity. The
integrated map (left) shows that the electron density ranges from
$\sim 1.5\times10^3$ cm$^{-3}$ at knot E (at $\sim$ 26 arcsec from
the outflow source) up to $5\times10^3$ cm $^{-3}$ in knot F (at
$\sim 2\arcsec5$ from knot E; see Figure \ref{fig9}).  It then drops
to $\sim 2\times10^3$ cm$^{-3}$ between knots F and H.  In knot H,
the electron density has a strong peak of $n_e\sim 3.5\times10^3$
cm$^{-3}$.

Interestingly, the electron density maps of the three velocity
channels show systematically increasing densities for decreasing
radial velocities. This is can be clearly seen at the positions of
knots E, F and J. Also, an increase in $n_e$ can be seen in the
region between knots F and G when going from the MVC to the LVC
electron density maps (see the two frames on the right of Figure
9).

\subsection{Knot E}

We now focus on the interesting structure of knot E (at $\approx
26''$ from the source, see Figures 3-7).  The structures observed
in the images and channel maps of this knot are shown in Figure 10.

% Figure 9: [SII] 6716/6731

We see that in the H$\alpha$ and [N II] images (2nd and 3rd
rows of the far left column of
Figure 10) knot E shows two side-by-side peaks. These peaks have
been observed in previous ground based (Reipurth et al. 1992) and
HST (Reipurth et al. 1997) H$\alpha$ images of HH~111. Most
interestingly, the two side-by-side peaks are well separated in the
LVC H$\alpha$ and [N II] maps (right column of Figure 10), approach
each other in the MVC map, and merge into a single, central peak
in the HVC map. Therefore, we find that the two side-by-side peaks
(previously observed in H$\alpha$ images, see Reipurth et al. 1997)
actually enclose a fainter, higher radial velocity emission region.

In the lower excitation [S II] and [O I] lines, knot E shows two
low intensity side-by-side peaks only in the LVC maps (1st and
4th rows of the far right column
of Figure 10), and these peaks are not present in the MVC and HVC
maps. The images of knot E in these lines (1st and 4th rows
of the far left column of Figure
10) show a conical structure which resembles the MVC emission.

\section{Summary}

We have obtained IFU spectra of the HH~111 jet at distances of
$24\to 49''$ from the outflow source. From the spectra, we obtain
position-velocity cubes for the [S II] 6716, 6731; H$\alpha$; [N
II] 6548, 6583 and [O I] 6300, 6360 emission lines.

We find a number of interesting results:

\begin{itemize}

\item we find that for all emission lines we have an increase in
the observed width of the jet for decreasing radial velocities.
This effect can be seen directly in the velocity channel maps
(Figures 3-6), in subtractions of pairs of velocity channel maps
(Figure 7) or more quantitatively in the widths measured for the jet
knots (Figure 8). A broadening in the HH~111 jet for decreasing
velocities has been previously observed by Riera et al. (2001), and
a qualitatively similar effect is seen in the CO emission (but at
widths in excess of $\sim 10''$, see Lefloch et al. 2007),

\item we find that the HH 111 jet is narrower in the lower ([S~II]
and [O~I]) than in the higher (H$\alpha$ and [N~II]) excitation
lines at the vicinity of some knots (namely, knots E, F K1 and
L). This effect has been previously seen in the H$\alpha$ and
[S~II] HST images of Reipurth et al. (1997). These previous
observations suggested that we might be seeing a sheath of non-radiative
shocks (seen only in H$\alpha$) surrounding the HH~111 jet beam,
but the detection of broad [N II] emission indicates that the shocks
in the sheath are indeed radiative,

\item from the [S~II] 6716/6731 line ratio we systematically find
larger electron densities at lower radial velocities (see Figure
9),

\item the ``twin peak'' structure of knot E is clearly seen in the
H$\alpha$ and [N II] images, but these two peaks merge into a central
peak in the [O I] and [S II] images. In all of the lines, the twin
peaks are seen at low radial velocities (see the LVC maps of Figure
10), but they approach and merge into a single peak at higher radial
velocities.

\end{itemize}

These characteristics can be qualitatively explained in terms of
an ``internal working surface'' model, in which the knots are
produced as a result of an outflow velocity time-variability. In
such a model, the variability produces pairs of shocks that travel
down the jet beam, producing a high pressure region that ejects
material into the cocoon of the jet. The observations of larger
widths for the higher excitation emission would then be interpreted
as higher velocity shocks produced by the material in this sideways
ejection against a slower moving cocoon. The low excitation emission
would be interpreted as coming from lower velocity shocks within
the jet beam (produced by the outflow velocity variability).

% Figure 10: field 8

\begin{figure*}
\centerline{
\includegraphics[scale=0.28]{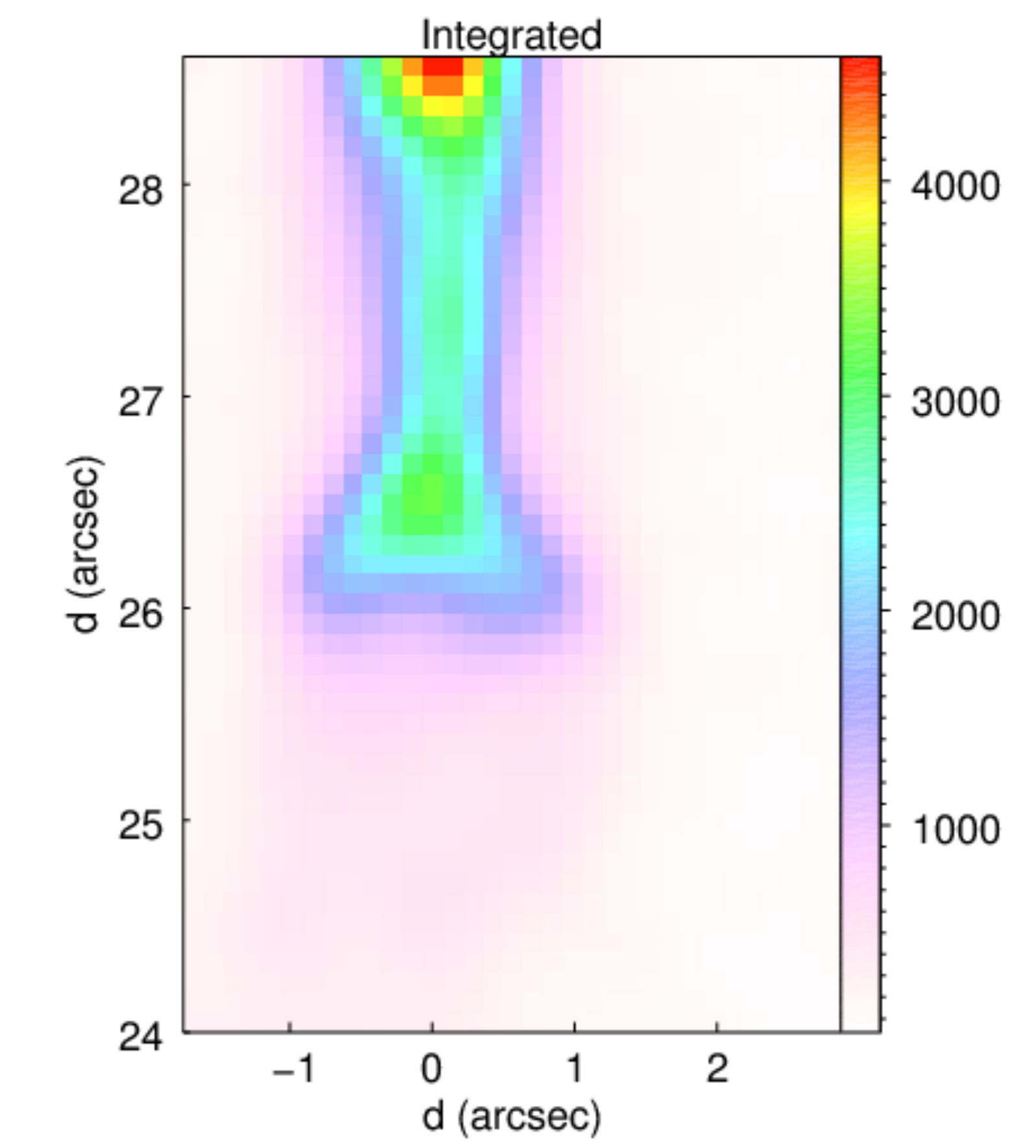}
\includegraphics[scale=0.255]{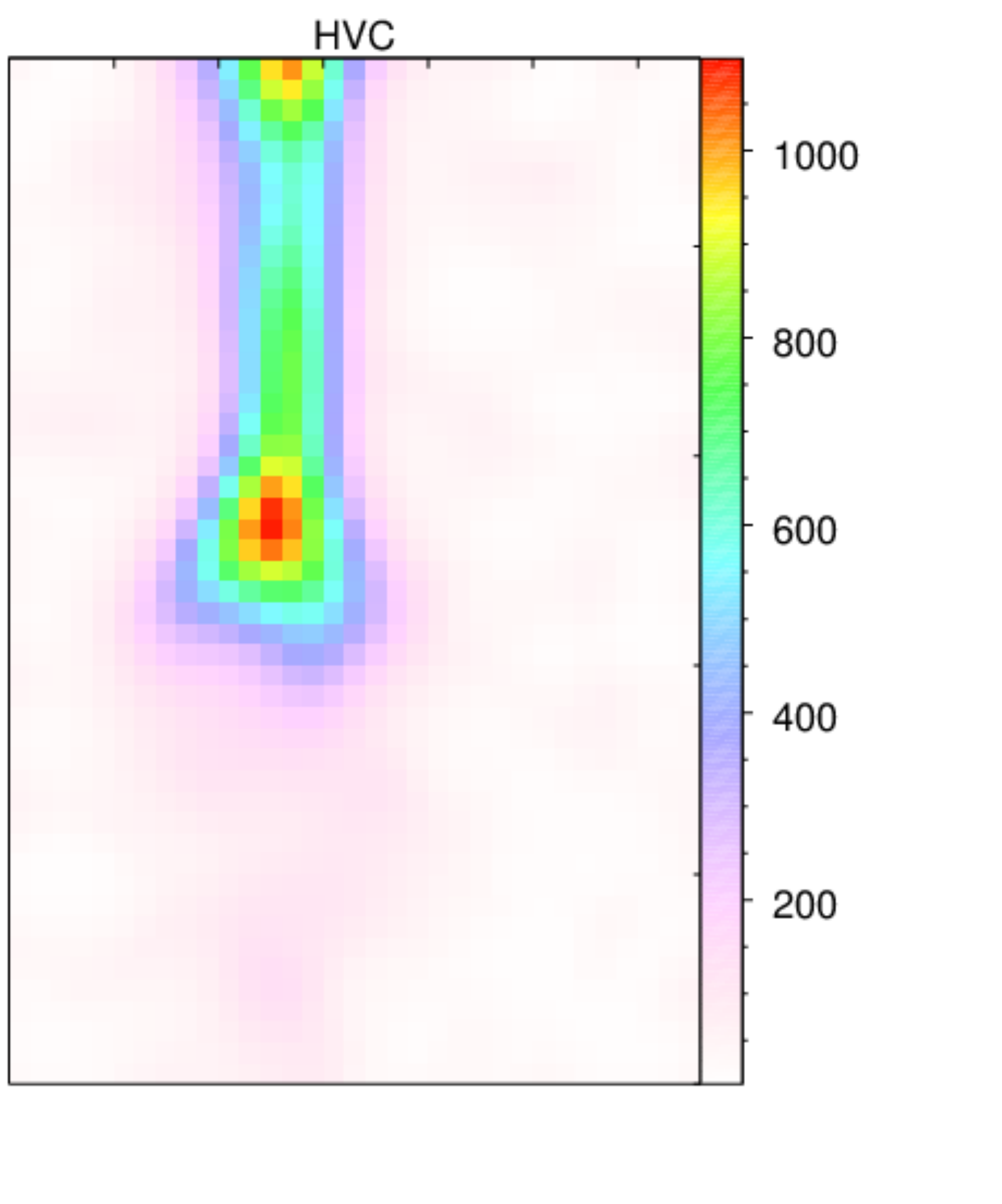}
\includegraphics[scale=0.255]{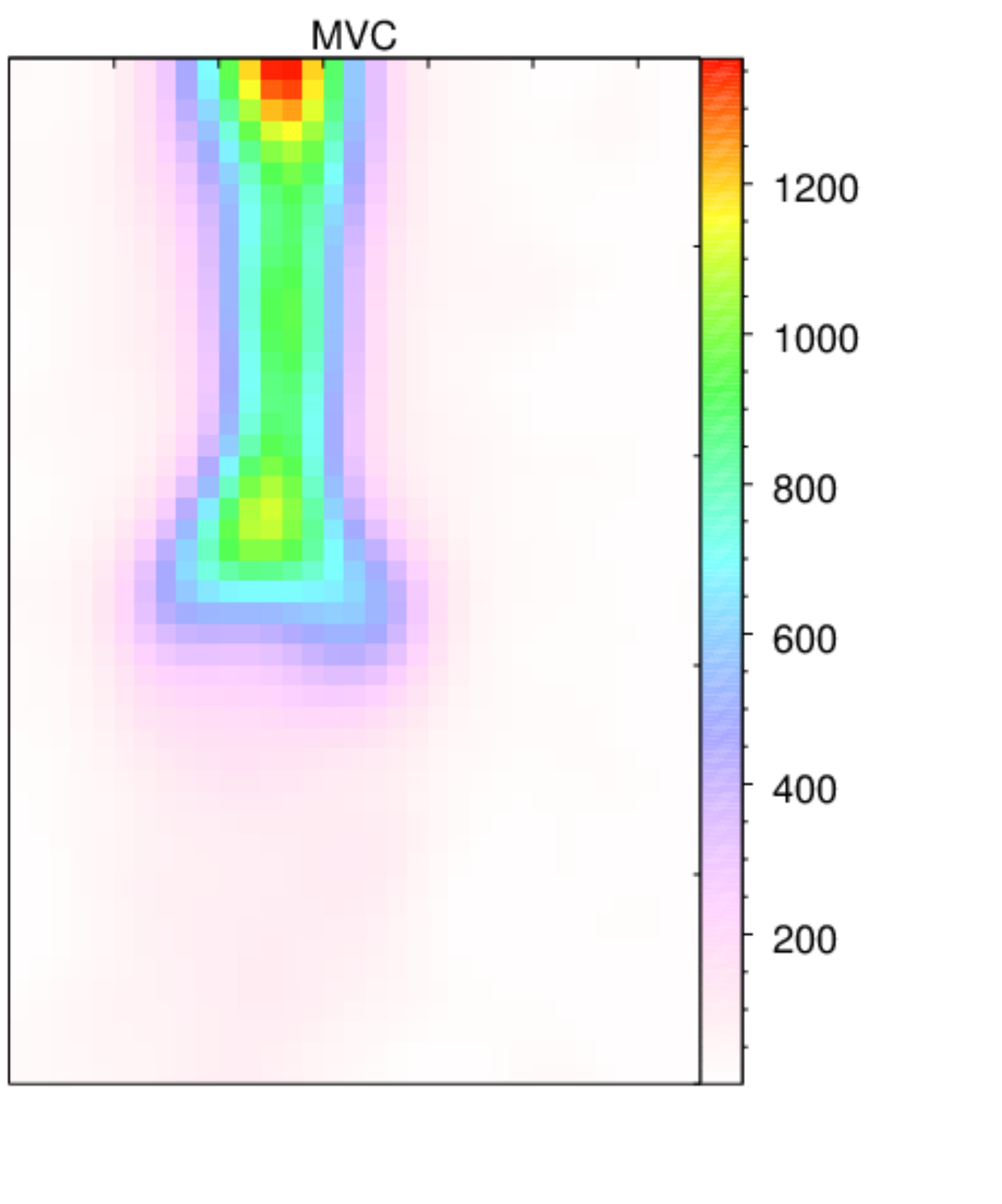}
\includegraphics[scale=0.255]{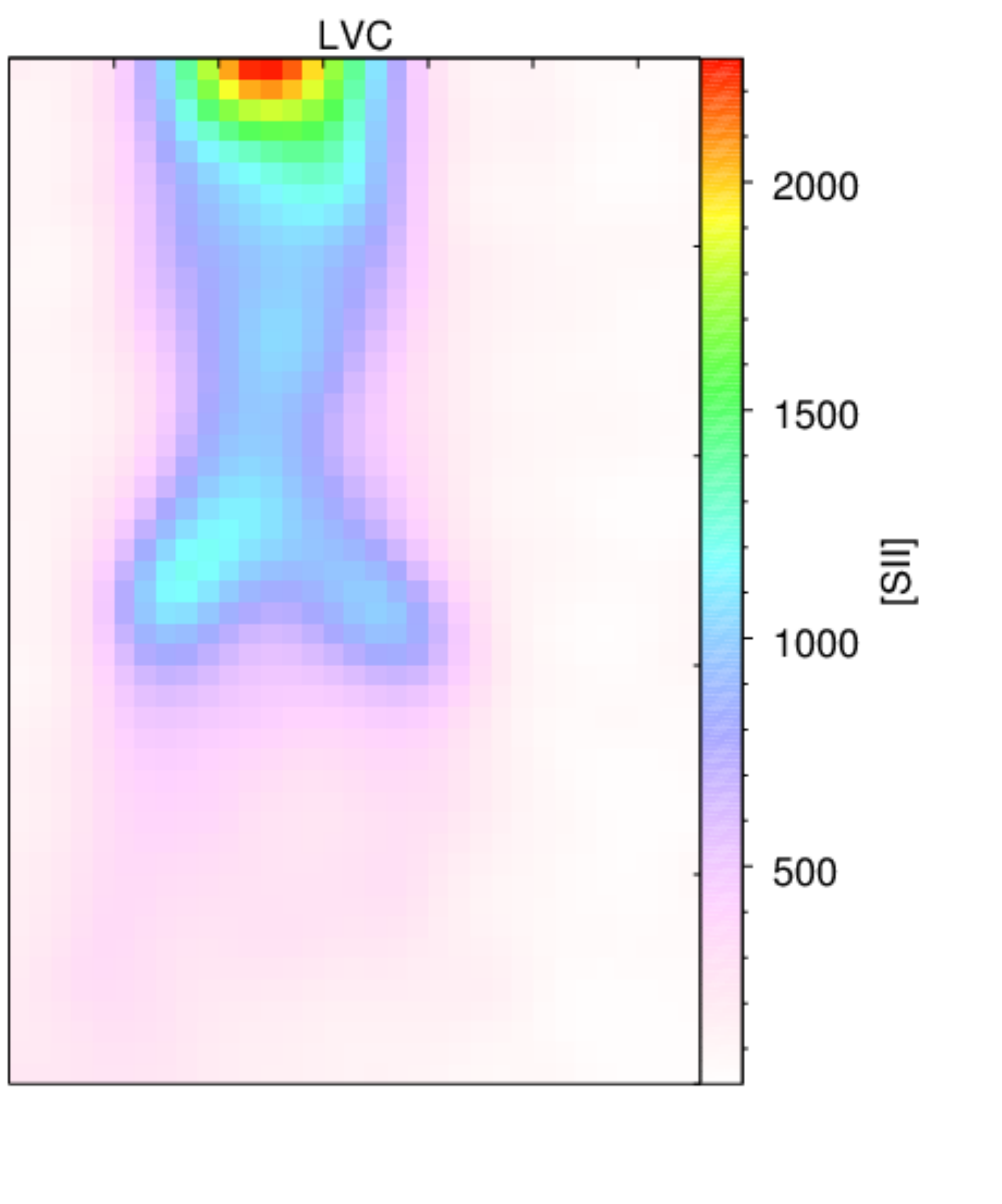}
}
\centerline{
\includegraphics[scale=0.28]{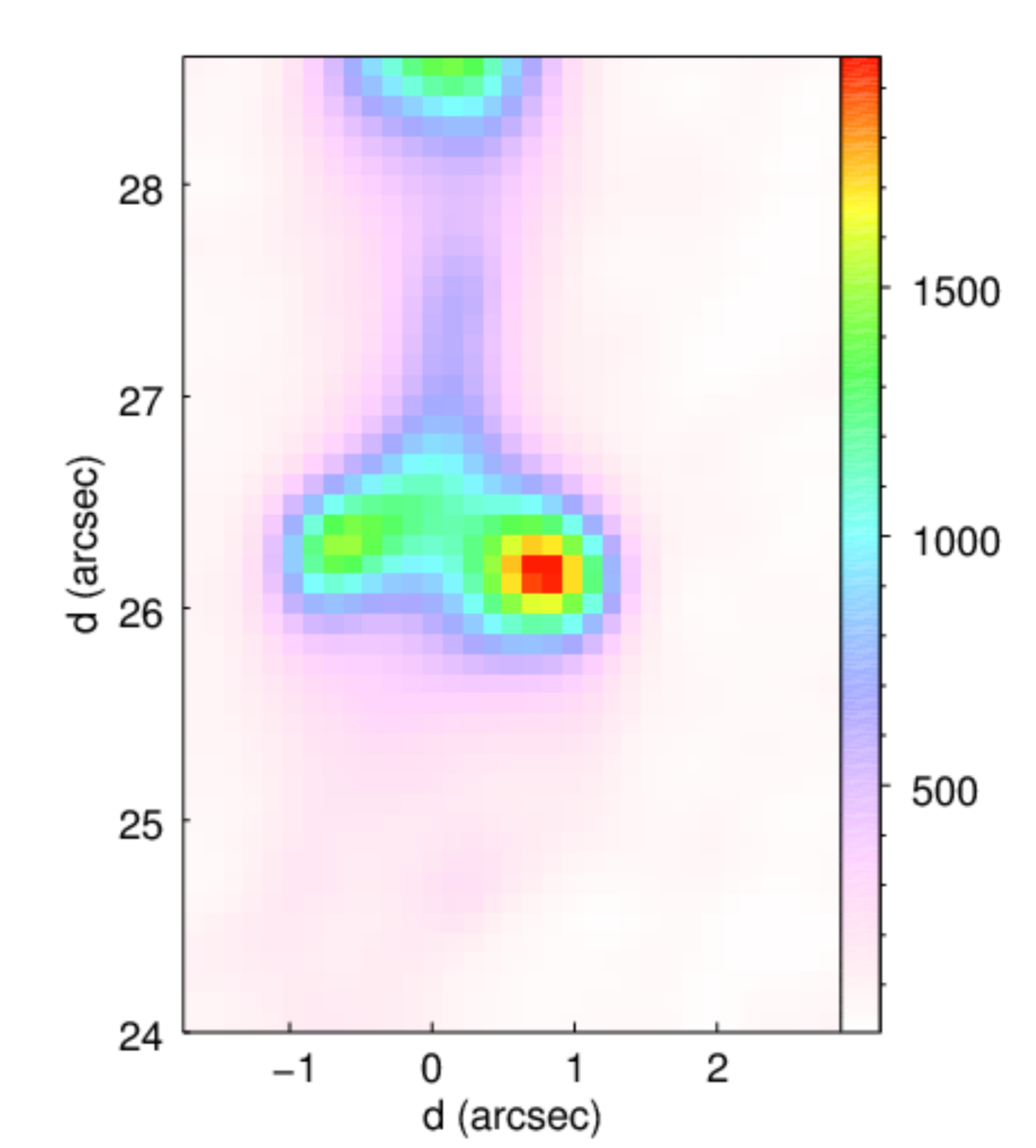}
\includegraphics[scale=0.255]{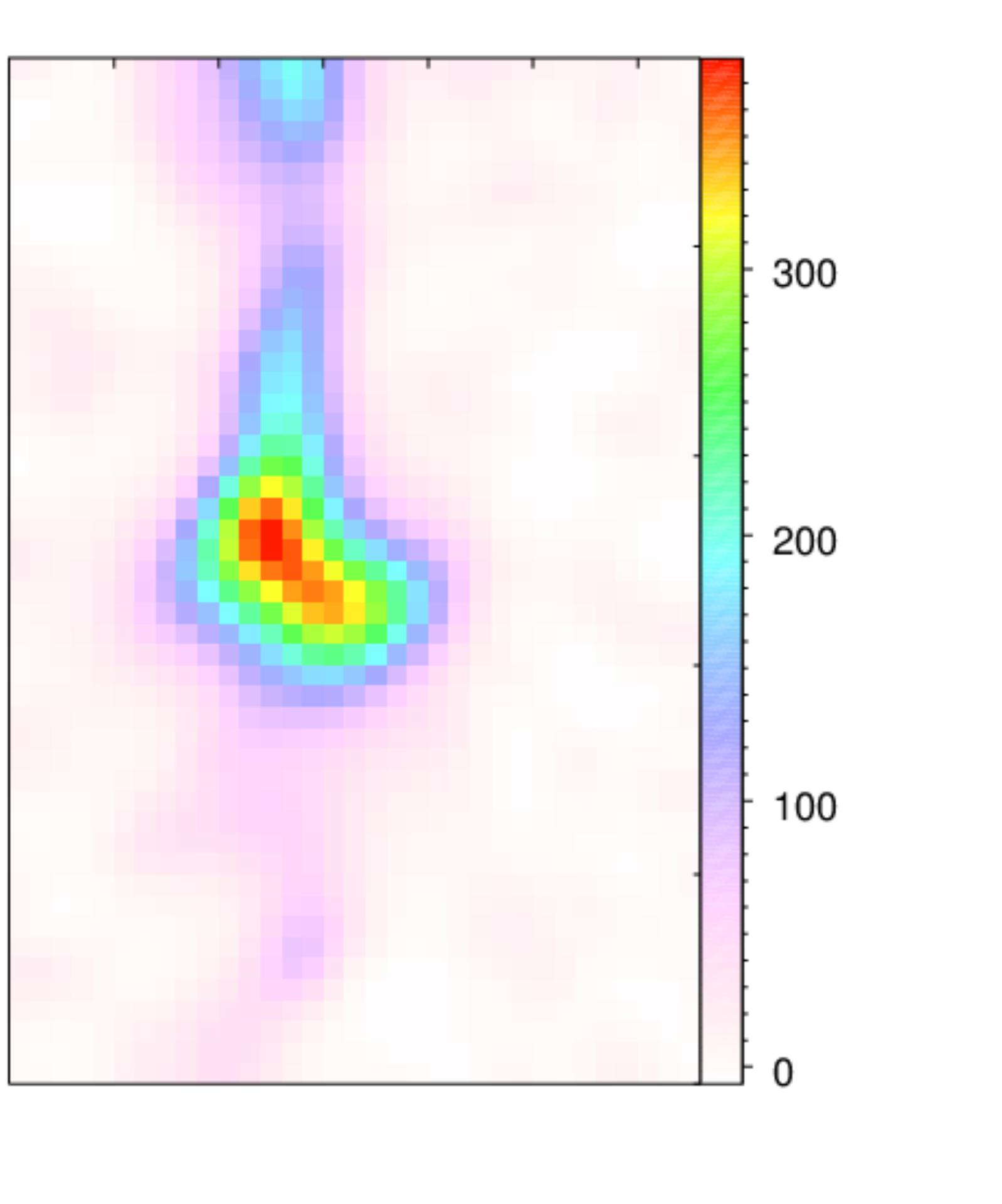}
\includegraphics[scale=0.255]{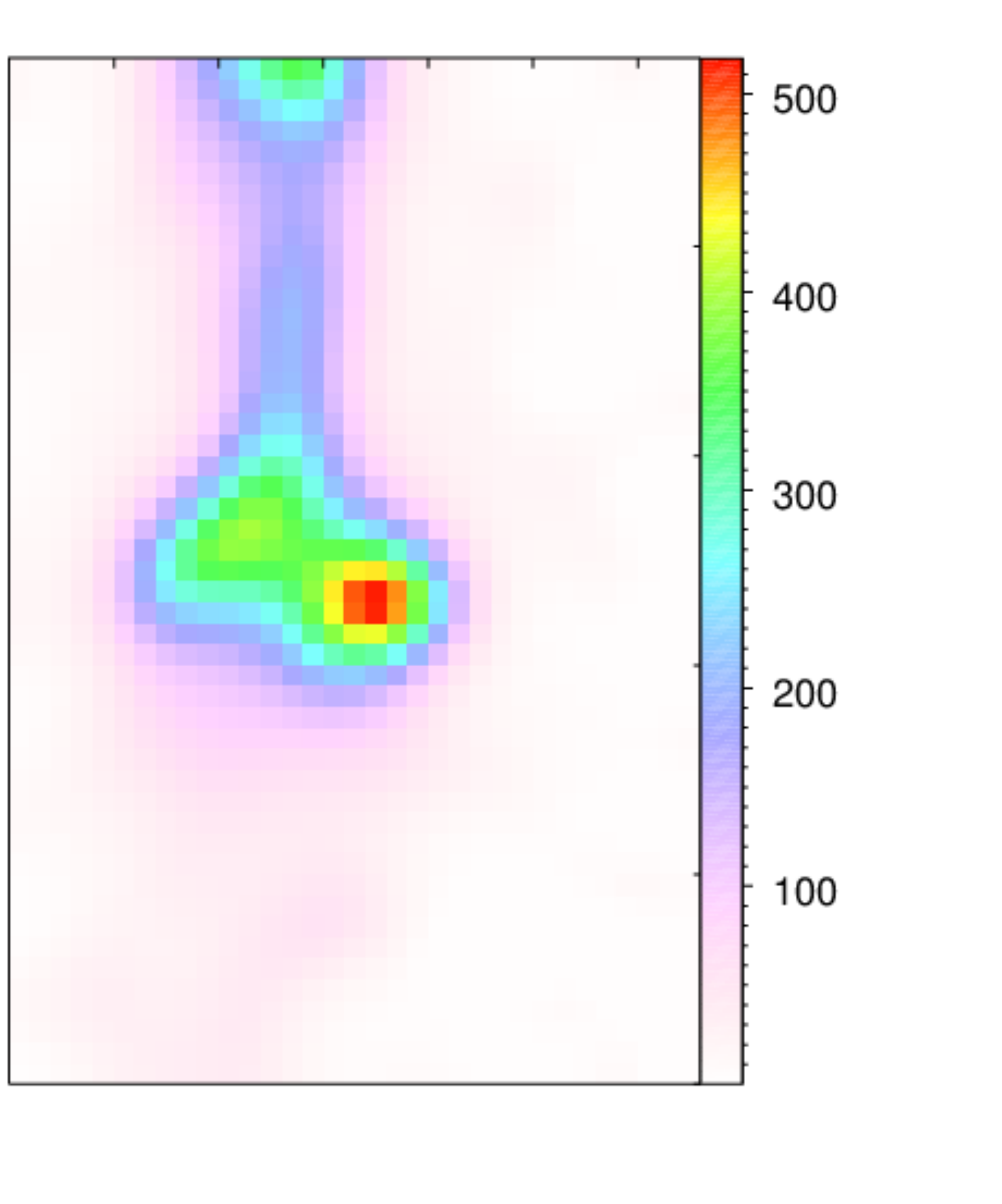}
\includegraphics[scale=0.255]{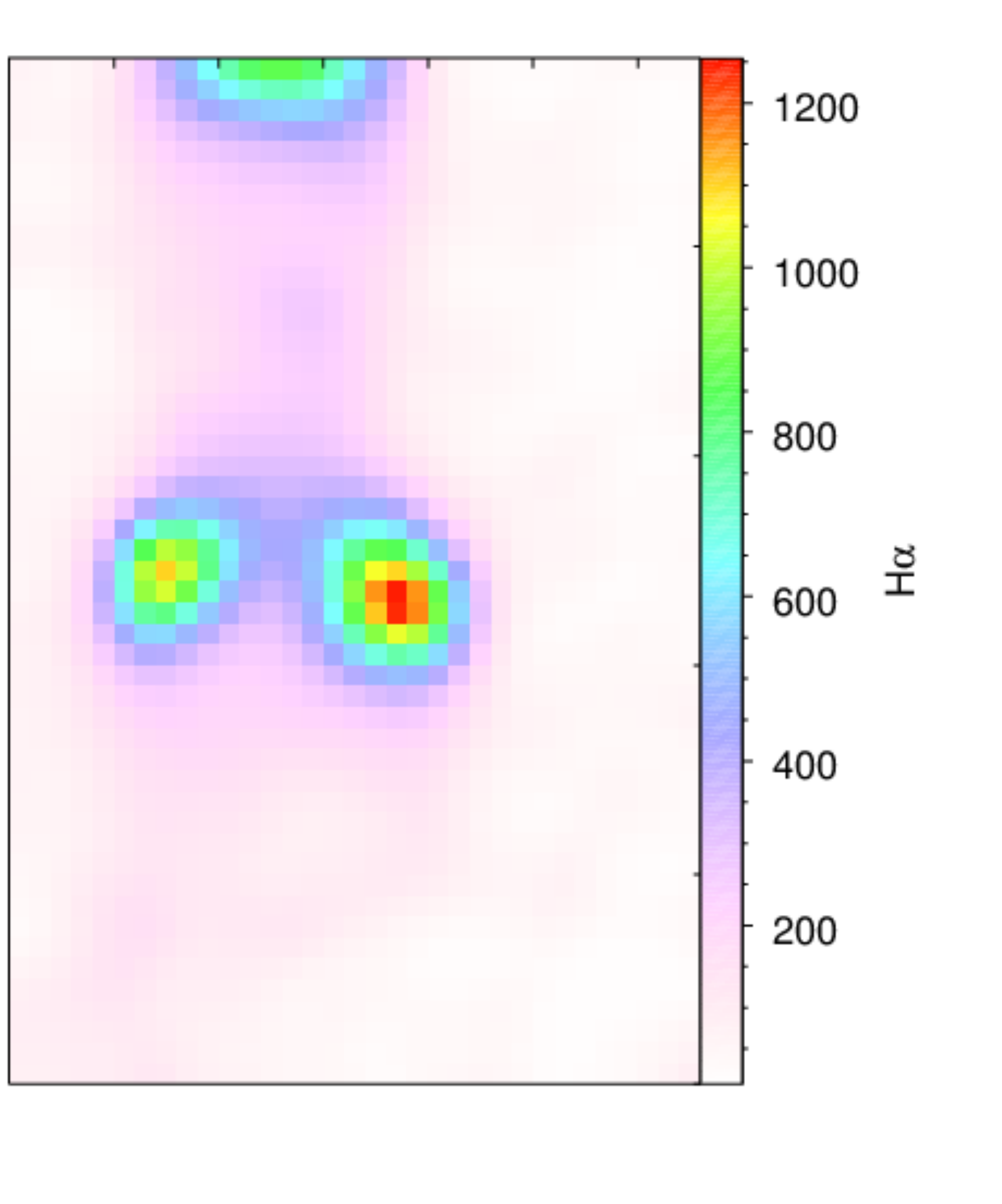}
}
\centerline{
\includegraphics[scale=0.28]{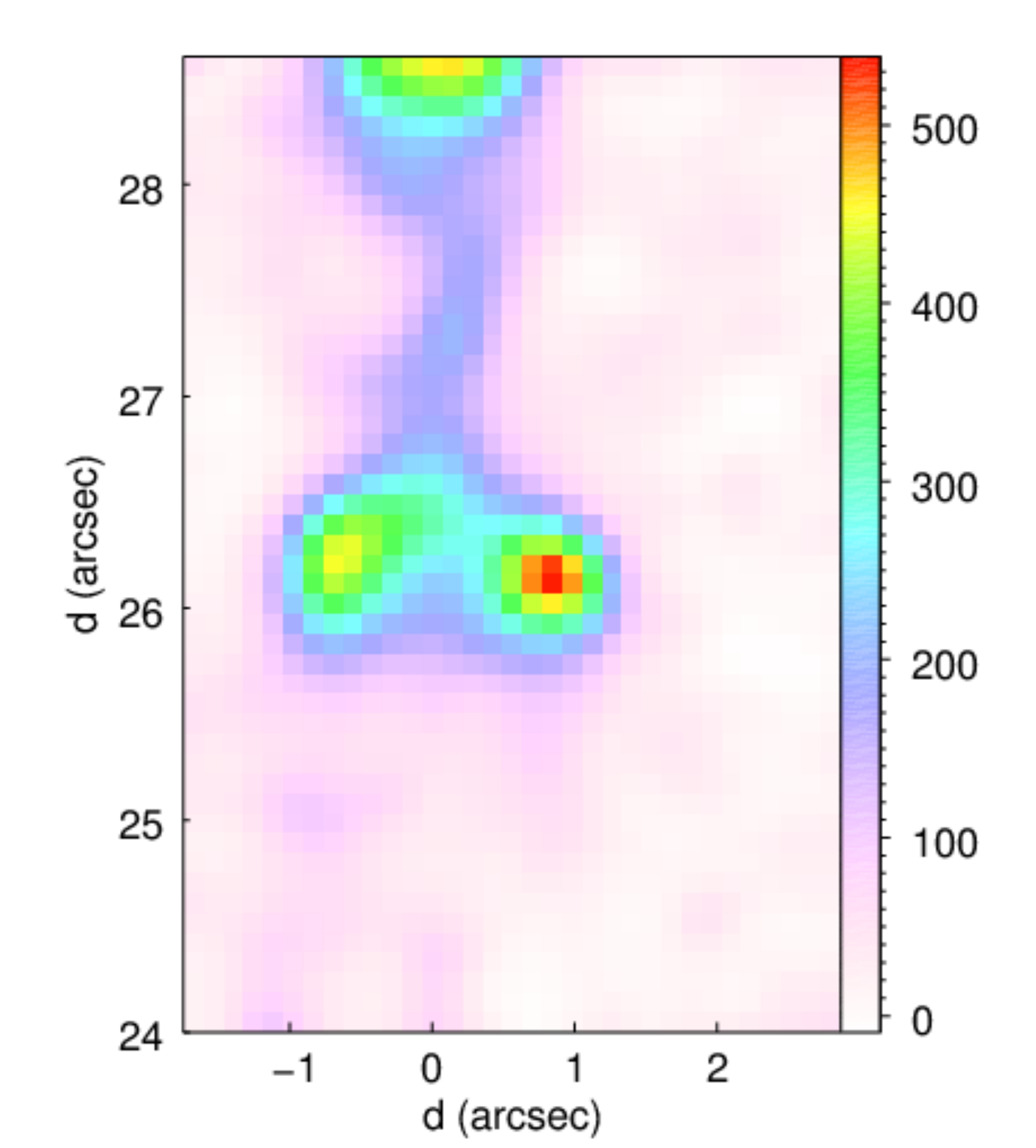}
\includegraphics[scale=0.255]{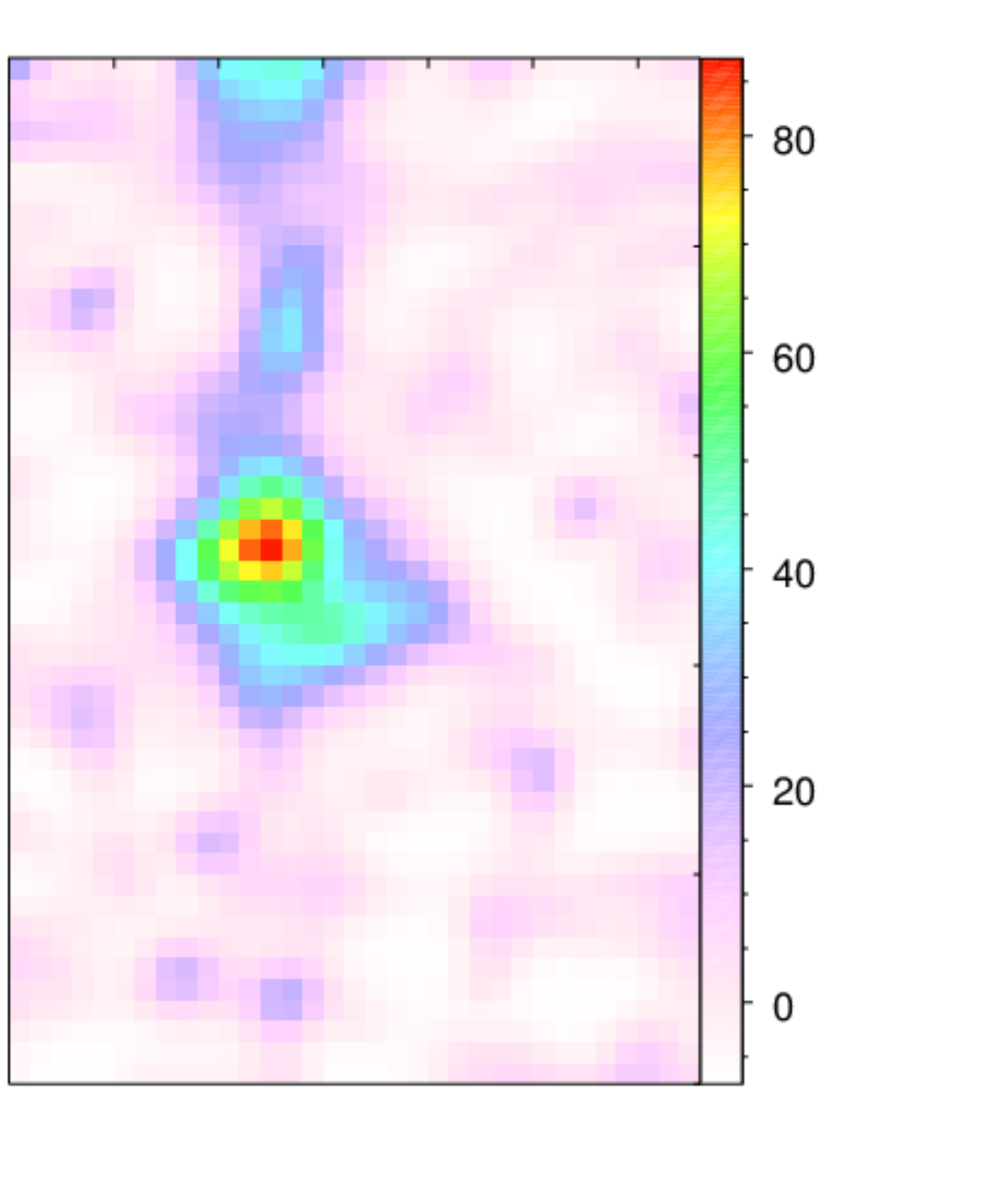}
\includegraphics[scale=0.255]{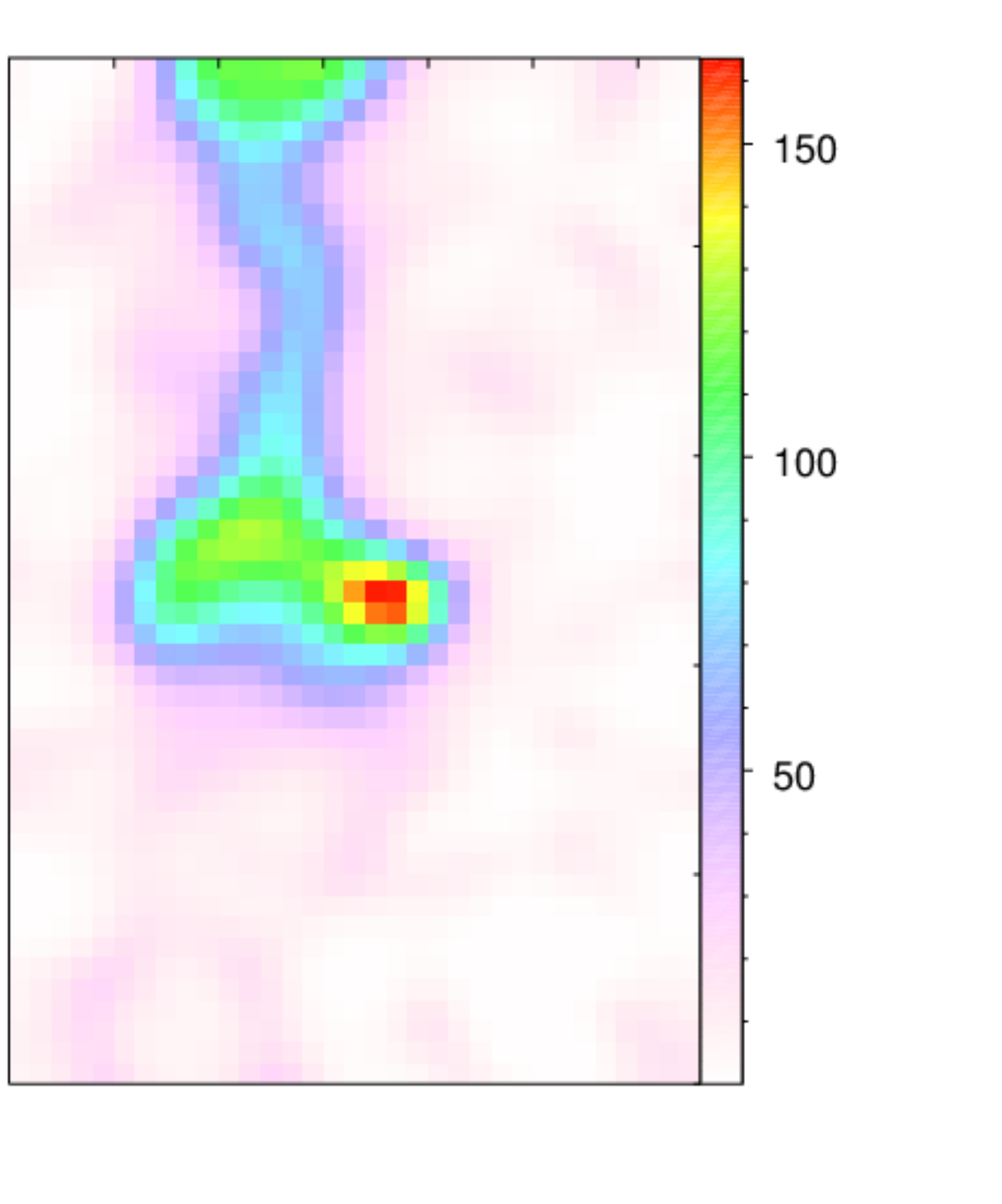}
\includegraphics[scale=0.255]{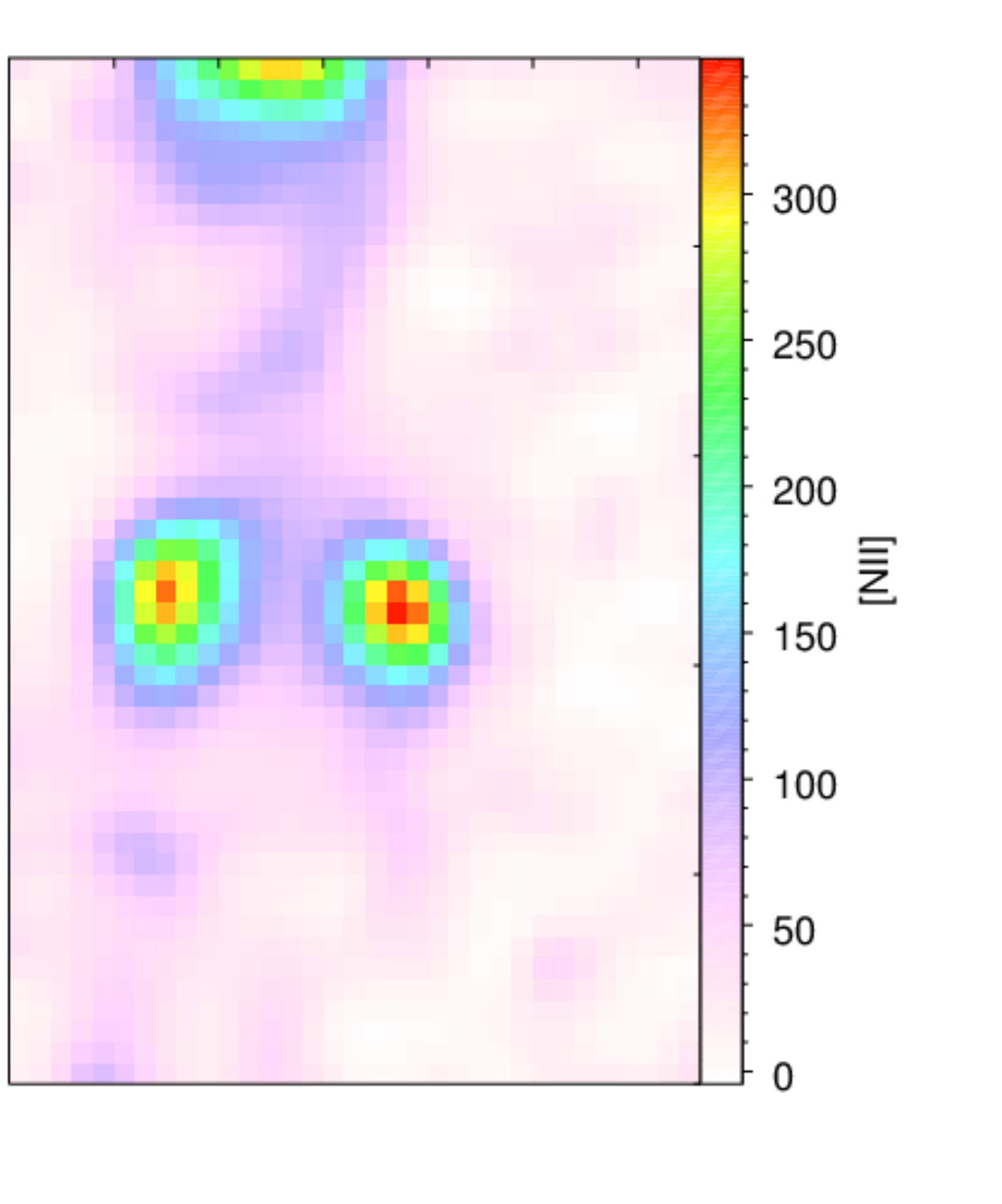}
}
\centerline{
\includegraphics[scale=0.28]{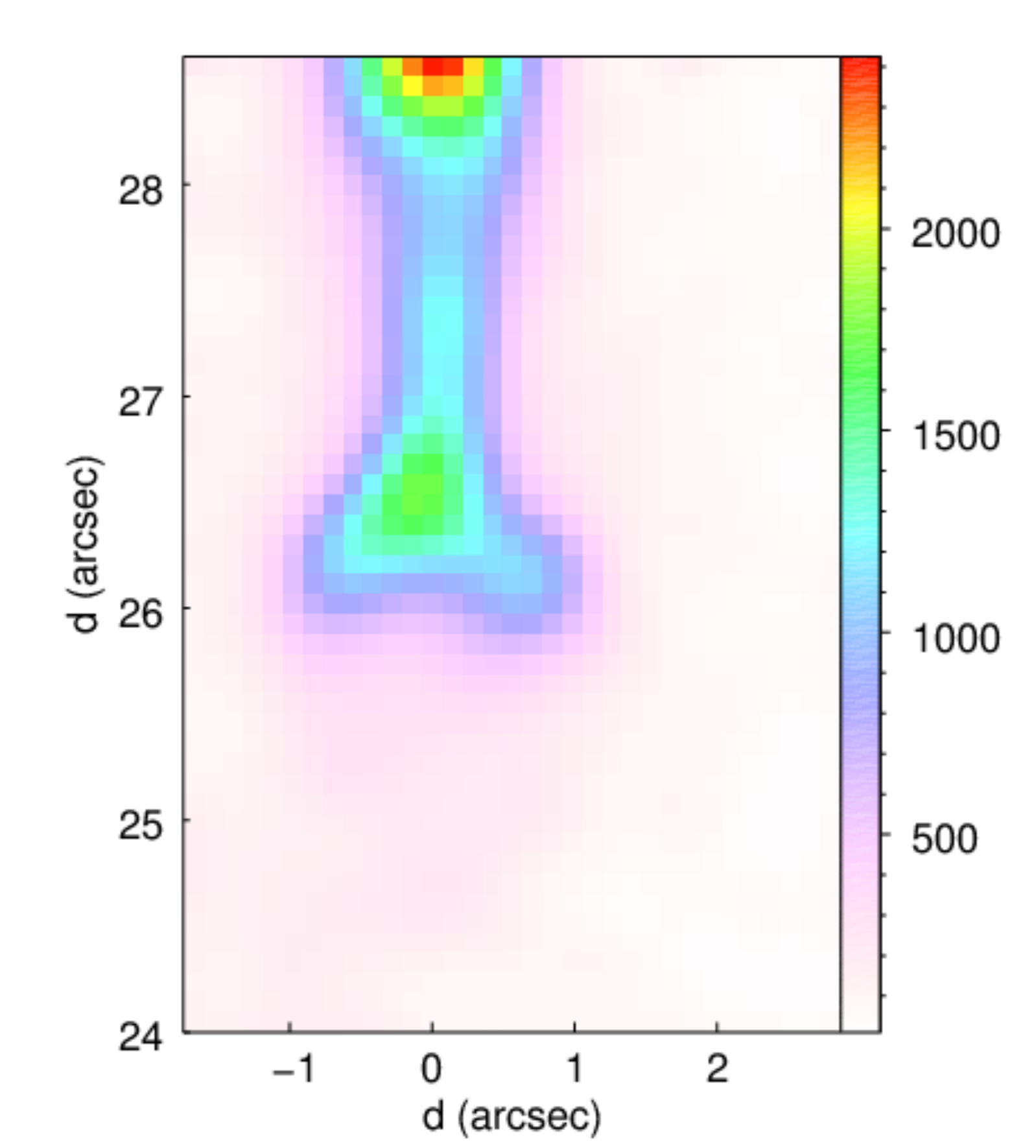}
\includegraphics[scale=0.255]{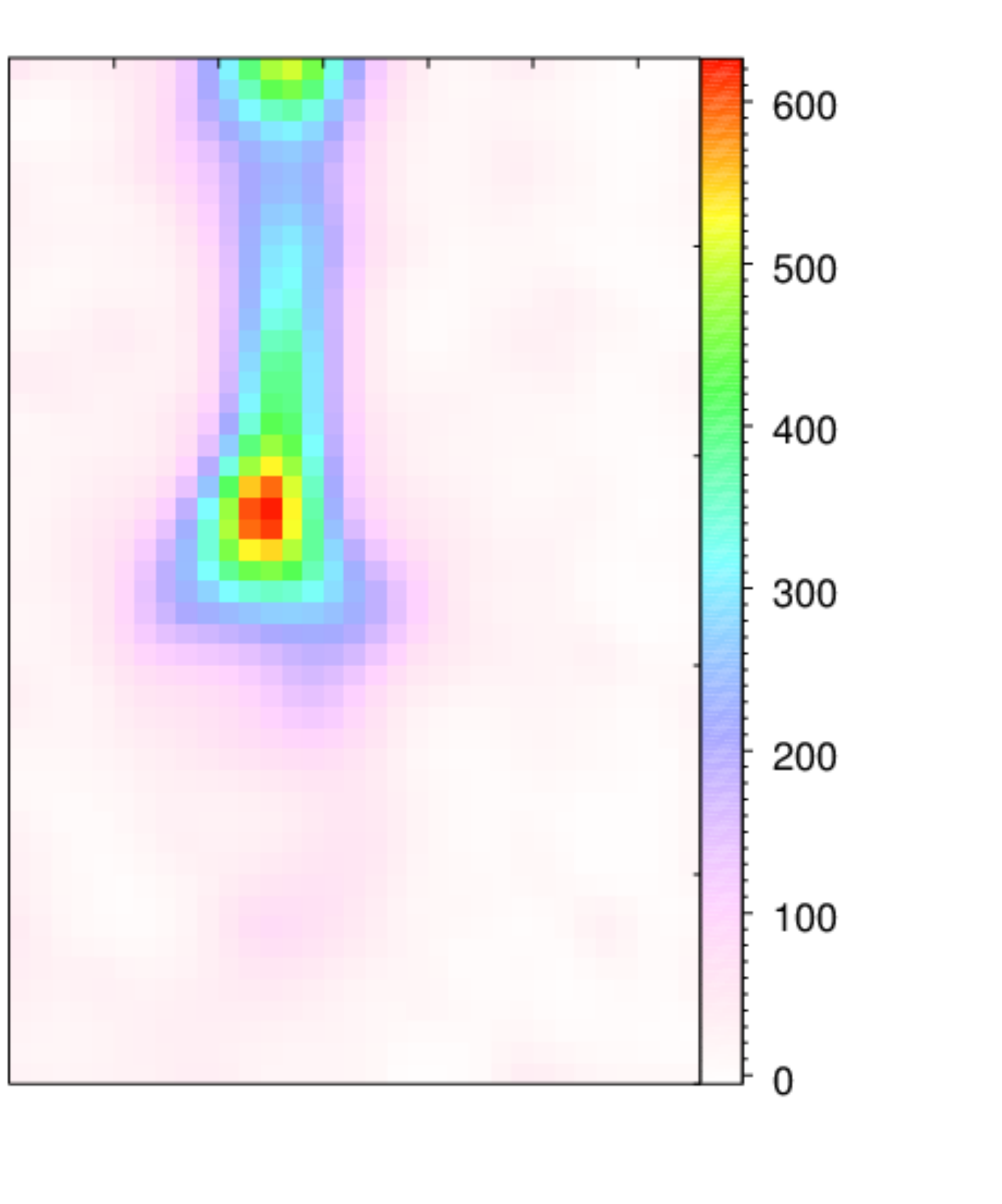}
\includegraphics[scale=0.255]{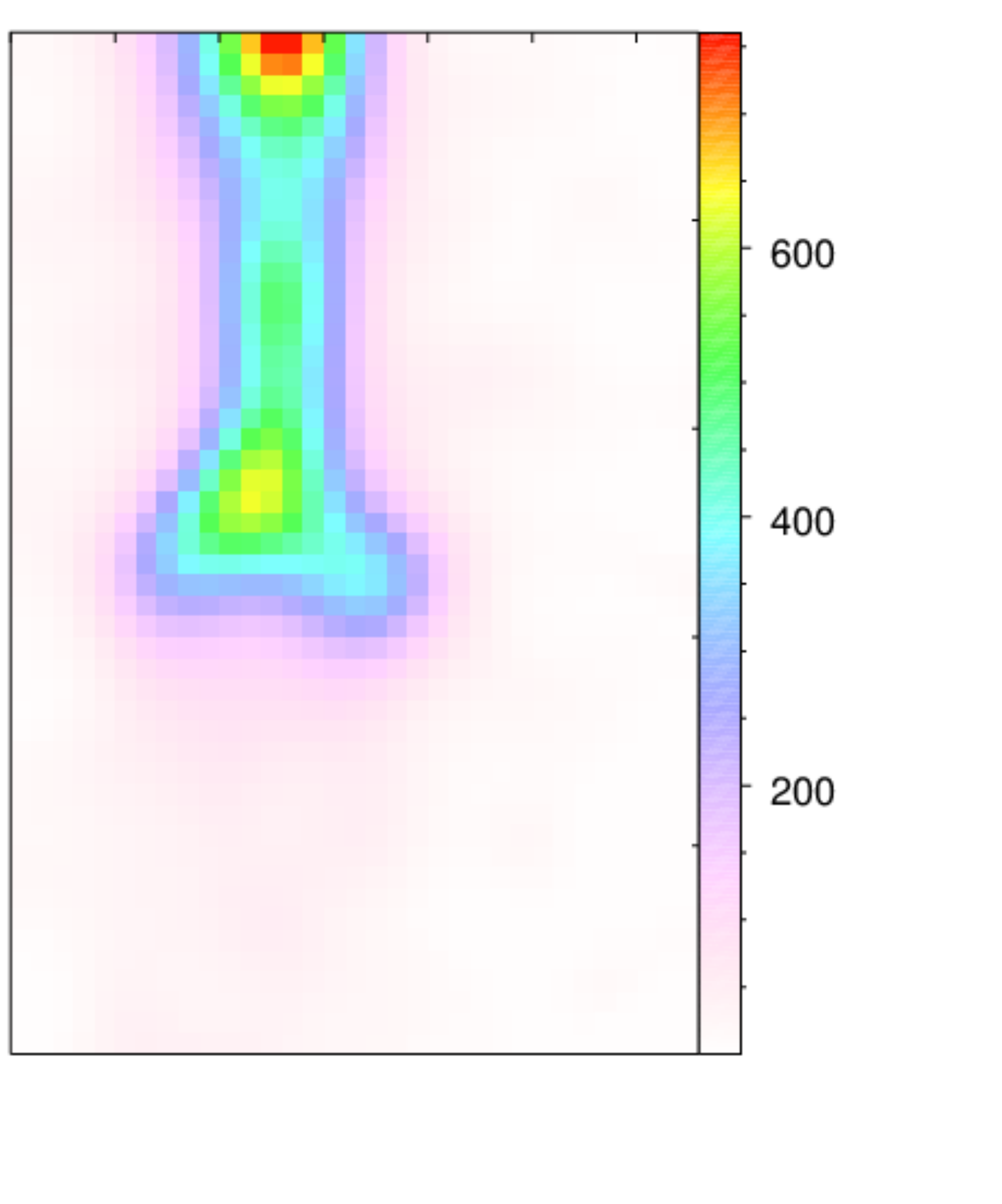}
\includegraphics[scale=0.255]{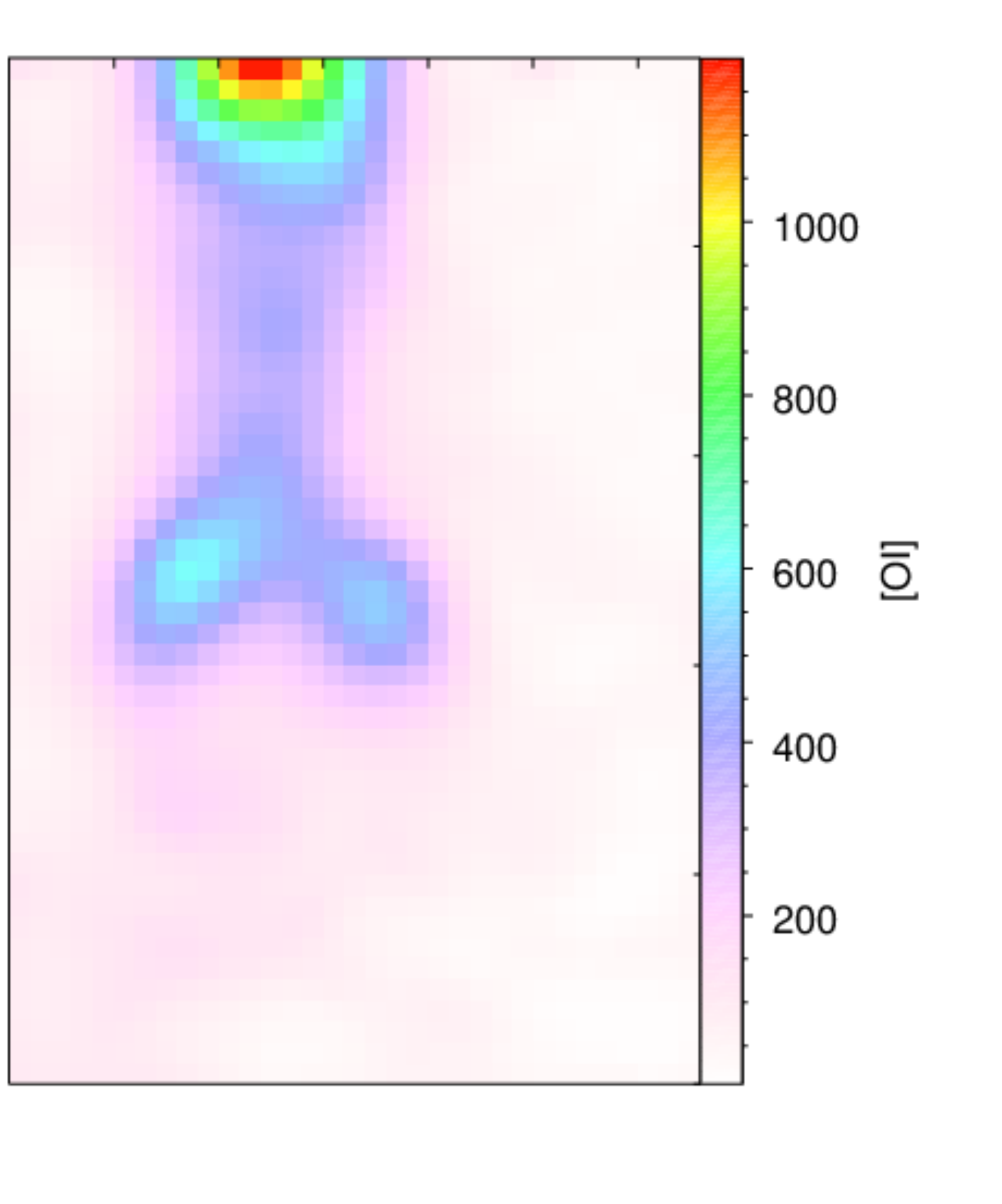}
}
\caption{{\it From left to right:} Integrated emission, HVC, MVC
and LVC for (from top to bottom) [S II] , H$\alpha$, [N II] and
[O I] for field 8 which contains the knot E. The axes
are in arcsec and indicates the distance from
the driving source. The maps are in arbitrary units (as
seen at the colorbars).}\label{fig10}
\end{figure*}

The shock structure of internal working surfaces was described by
Raga et al. (1990), and numerical simulations have been done for
reproducing H$\alpha$ images (Raga et al. 2002b) and long-slit
spectra (Masciadri et al. 2002) of the HH~111 jet. The data presented
here clearly provide a substantial increase in the observational
constraints, which should be a challenging test for future models
of jet knots produced by an outflow variability or by other mechanisms
(see e.g., Micono et al. 1998). In particular, our data will give
interesting constraints on the nature of the cross section of the
outflow (Raga et al. 2011) or of the possible presence of two
distinct components (Te\c sileanu et al. 2014).

We can speculate that MHD effects might be present, and might
help to shape the morphology of the HH 111 emission knots as well
as their line profiles. It is worth to note that recent and
sophisticated MHD models predict that beam recollimation (by the
hoop stress) plays an important role in order to enhance the
brightness near the jet axis, as well as to accelerate the inner
jet beam, making it faster than the surrounding jet material. This
general result was obtained by Te\c sileanu et al (2014), Hansen,
Frank \& Hartigan (2014) and  Staff et al. (2015), in spite of the
differences in the initial setup in these works. In Hansen, Frank
\& Hartigan (2014), a pure toroidal magnetic field is used to study
the jet propagation, while the disk wind solution is incorporated
in the models from Staff et al (2014) and Te\c sileanu et al. (2014),
who actually introduced a two component (stellar jet plus disk wind)
flow.

Another feature shown by some HH jets (including some of the
knots studied here) is an increase of the FWHM of the jet beam with
distance from the driving source. This feature has been predicted
in the models presented by Te\c sileanu et al. (2014)  and compared
with available data from RW Aur, HH 30 and HL Tau, while the solutions
presented in Staff et al. (2014) were compared with data from DG
Tau, HN Tau, RW Aur and UZ Tau. In both cases, the distances covered
by the simulations were suitable for the study of the microjets
($\sim 100$ AU). Whether or not these effects are still important
farther away from the driving source (the knot E is at $\sim 10\,000$
AU from the VLA 1 source) is an issue that should be addressed
properly in future work.

\acknowledgments

We would like to thanks the referee for his/her suggestion.  We are
grateful to R. Carrasco  and B. Miller (from Gemini South telescope),
and R. Schiavon (Liverpool John Moores University) for help us with
data reduction and scripts. We are thankful to T. Ricci and J.
Steiner for enlightening discussions about the PCA technique and
its applications to data-cube, and for providing the pipeline for
data reduction. AHC, MJV and HP thanks Cnpq/CAPES for financial
support using the PROCAD project (552236/2011-0) and CAPES/CNPq
Science without Borders program, under grants 2168/13-8 (AHC) and
2565/13-7 (MJV). J. Feitosa was fully supported by a INCT-A
scholarship. AR acknowledges support from CONACyT grants 101356,
101975 and 167611 and the DGAPA-UNAM grants IN105312 and IG100214.
This paper has been submitted during our (AHC and MJV) 
sabbatical leave in Grenoble (IPAG/UJF), and we are very thankful
to J. Bouvier and J. Ferreira for their warm hospitality.

{\it Facilities:} \facility{Gemini (GMOS)}.

%%%%%%% Begining of the References ***********

% End of the References ***************

%\clearpage

\end{document}